\RequirePackage{latexrelease}
\documentclass{aa} 

\usepackage{graphicx}
\usepackage{tensor}
\usepackage{lscape}
\usepackage{txfonts}

\usepackage{amsmath}           
\usepackage{amsfonts}          
\usepackage{amssymb}
\usepackage[export]{adjustbox}           
\usepackage{fix-cm}
\usepackage{multirow}
\usepackage{units}

\makeatletter
\renewcommand*\aa@pageof{, page \thepage{} of \pageref*{LastPage}}
\makeatother

\usepackage{xcolor}
\usepackage{natbib,twoopt}
\usepackage[hyphenbreaks]{breakurl}
\usepackage[breaklinks]{hyperref}      
\bibpunct{(}{)}{;}{a}{}{,}             
\definecolor{cobalt}{rgb}{0.06, 0.2, 0.65}
\hypersetup{
  colorlinks,
  citecolor=cobalt,
  linkcolor=[rgb]{0.8, 0.2, 1.0},
  urlcolor=cobalt,
}
\makeatletter
  \newcommandtwoopt{\citeads}[3][][]{\href{http://adsabs.harvard.edu/abs/#3}%
    {\def\hyper@linkstart##1##2{}%
     \let\hyper@linkend\@empty\citealp[#1][#2]{#3}}}
  \newcommandtwoopt{\citepads}[3][][]{\href{http://adsabs.harvard.edu/abs/#3}%
    {\def\hyper@linkstart##1##2{}%
     \let\hyper@linkend\@empty\citep[#1][#2]{#3}}}
  \newcommandtwoopt{\citetads}[3][][]{\href{http://adsabs.harvard.edu/abs/#3}%
    {\def\hyper@linkstart##1##2{}%
     \let\hyper@linkend\@empty\citet[#1][#2]{#3}}}
  \newcommandtwoopt{\citeyearads}[3][][]%
    {\href{http://adsabs.harvard.edu/abs/#3}
    {\def\hyper@linkstart##1##2{}%
     \let\hyper@linkend\@empty\citeyear[#1][#2]{#3}}}
\makeatother

\usepackage[font=small]{caption}

\newcommand{\sidebysidecaption}[4]{%
\captionsetup{justification=justified}
  \begin{minipage}[t]{#1}
    \vspace*{0pt}\captionsetup{width=\textwidth}
    #3
  \end{minipage}
  \hfill%
  \begin{minipage}[t]{#2}
    \vspace*{0pt}\captionsetup{width=\textwidth}
    #4
  \end{minipage}%
}%

\usepackage{orcidlink}

\begin{document}

\title{The magnetic filling in magnetically arrested accretion disk simulations and its impact on the jet in M87}

\author{Felix Glaser\inst{\ref{inst1}}\orcidlink{0009-0005-5870-4195}, Christian M. Fromm\inst{\ref{inst1},}\inst{\ref{inst2}}\orcidlink{0000-0002-1827-1656}, Yosuke Mizuno\inst{\ref{inst5},}\inst{\ref{inst6},}\inst{\ref{inst7}}\orcidlink{0000-0002-8131-6730}, Matthias Kadler\inst{\ref{inst1}}\orcidlink{0000-0001-5606-6154}, Karl Mannheim\inst{\ref{inst1}}\orcidlink{0000-0002-2950-6641}}
\institute{Institute for Physics and Astronomy, University of Würzburg, Emil-Hilb-Weg 31, 97074 Würzburg \label{inst1} \\
    \email{felix.glaser@stud-mail.uni-wuerzburg.de, christian.fromm@uni-wuerzburg.de}
    \and Institute for Theoretical Physics, Goethe Universit\"at Frankfurt, Max-von-Laue-Str. 1, 60438 Frankfurt, Germany \label{inst2}
    \and Tsung-Dao Lee Institute, Shanghai Jiao Tong University, Shanghai, 201210, People’s Republic of China \label{inst5} \and School of Physics and Astronomy, Shanghai Jiao Tong University, Shanghai, 200240, People’s Republic of China
    \label{inst6} \and  Key Laboratory for Particle Astrophysics and Cosmology (MOE) and Shanghai Key Laboratory for Particle Physics and Cosmology, Shanghai Jiao Tong University, Shanghai 200240, People’s Republic of China \label{inst7}\\
}

\date{Received 20 May 2025 / Accepted 14 February 2026}
\abstract
{Magnetically arrested accretion disks (MADs) in black-hole (BH) jet-launching simulations are very successful in modeling low-luminosity active galactic nuclei (AGNs) such as M87*. The Fishbone-Moncrief (FM) torus is well established for this purpose in numerical astrophysics. The extent of the magnetic vector potential inside the FM torus --which we dub the filling factor-- has not been studied before in the case of MAD simulations.}
{In detail, we stress the impact on the jet morphology, efficiency, and linear polarization imprints of the filling factor as a significant initial parameter of these simulations.}
{We employed five 3D, general-relativistic magnetohydrodynamic (GRMHD) simulations initialized with large-scale tori, which are immersed in weak, poloidal magnetic fields. To study the impact of the spatial extent of the initial magnetic field, and hence the magnetic energy content in the torus, we scaled it with the filling factor with regard to the poloidal geometric area of the mass-density distribution. A common choice of the filling factor is complimented and investigated in terms of altered energetics and angular-momentum transport. Further, we investigated the polarized, radiative imprints of synchrotron emission on M87 at 86 GHz, comparing them with very long baseline interferometry (VLBI) observations. Therefore, we performed general-relativistic, polarized radiative transfer calculations on the GRMHD data, modeling thermal and nonthermal electron distributions.}
{Our simulations show that elevated filling factors significantly increase the electromagnetic (EM) energy contributions and outward angular-momentum transport in the jet due to the initially increased magnetic energy content in the torus. High magnetic fillings exhibit increased linear polarization fractions, agreeing with the observed $\sim 15\%$ in M87*. We find the jet morphology more prone to disk vertical flux tubes generated by MAD events. We show that GRMHD simulations bracket the jet width measurements at the jet base in M87*.}
{Increased magnetic filling of the FM torus produces jets that are noticeably brighter downstream compared to our reference models; hence, we also find high fillings well suited for extended GRMHD jet models of other low-luminosity AGNs.}


\authorrunning{Glaser et al.}
\titlerunning{The magnetic filling in magnetically arrested accretion disk simulations and its impact on the jet in M87}
\maketitle

\section{Introduction}
Messier 87 (M87), a giant elliptical galaxy in the Virgo cluster, has provided an important laboratory in the field of extragalactic jet astrophysics for many decades. The jet, originating from a compact radio core in the galaxy center, where the active galactic nucleus (AGN) is located, has been studied extensively across the electromagnetic spectrum, ranging from the radio regime up to $\gamma$-rays \cite[e.g.][]{Reid_1982,  Hada_2013, Kim2018, Snios_2019, magic_2020, Lu_2023}. The Event Horizon Telescope Collaboration (EHTC; \cite[][]{EHT1}) resolved the radio core of M87* at $1.3\,\unit{mm}$ employing the very long baseline interferometry (VLBI) technique and interpreted the ring-like structure as the gravitationally lensed synchrotron emission, originating from the accretion disk of a supermassive black hole (SMBH). In more recent observations at $86\,\unit{GHz}$, \cite{Lu_2023} revealed that the jet and the radio core are indeed linked, which was interpreted as the jet base connecting to the accretion flow of relativistic plasma. Super-resolution images of mentioned observations with the Global Millimetre VLBI Array (GMVA) and Very Long Baseline Array (VLBA) at $43\,\unit{GHz}$ were obtained, employing the \texttt{resolve} algorithm \cite[see][]{kim2024,Kim_2024}. The polarimetric EHT observations obtained in 2017 provide viable information on the dynamically important magnetic-field structure. From these observations, a linear polarization fraction of $\sim 15\,\unit{\%}$ \cite[][]{EHTVII} and an upper bound on the circularly polarized emission fraction of $\langle|\nu|\rangle < 3.7\%$ \cite[][]{EHTIX} was found. These results constrain theoretical models such as general-relativistic magnetohydrodynamic (GRMHD) simulations of the plasma flow in the vicinity of the black hole (BH). These simulations can be compared to observations by means of general-relativistic radiative-transfer (GRRT) post-processing calculations on the GRMHD data. \cite{Yang_2024} found the 86 GHz and 43 GHz jet morphology of similar GRMHD simulations to fit the observations very well. \cite{Cruz_Osorio_2022} and \cite{Fromm2022} found high spin parameters of $a\sim0.5-0.94$ to be favorable to match the broad-band spectrum and jet morphology of M87* in wide parameter-space studies of GRMHD models. In this work, we focused on the comparison of five 3D GRMHD models and their impact on the electromagnetic footprint of the M87 jet. All share similar initial conditions, but differ with respect to their initial magnetic-field configuration. We used the Fishbone-Moncrief (FM) torus \cite[][]{1976FishboneMoncrief} and initialized it with a weak poloidal magnetic field. In most studies, the poloidal extent of the initial magnetic field is set with the offset of the initial vector potential from a zero cutoff to $0.80$. We compared this model with four simulations that fill the initial torus with varying spatial extents of a weak poloidal magnetic field. These models exhibit drastically altered energetic-flux, angular-momentum transport properties and jet-generation efficiencies. \cite{Chatterjee_2022} studied the behavior and drivers of angular-momentum transport in magnetically arrested disk (MAD) simulations compared to standard- and normal-evolution (SANE) simulations in non-spinning BH systems to isolate the side effects of frame dragging. We complemented this study with the case of BHs with high spin and compared the impact of the magnetic filling of the torus. Recent studies by \cite{jacqueminide2025} complement explanations for the underlying mechanism of MAD states by studying the inward advective and outward diffusive cycles these simulations exhibit, and, based on that, they derived a first analytical expression for the recurrence timescale of MAD states. Further parametrical investigation of the MAD setup by \cite{cho2025} found that the overall variability in the simulations is sensitive to the ratio of rotational and magnetic energy in the initial setup, which connects well to our study, where we tuned the initial magnetic-energy content. \\ \noindent
This paper contains two major parts. In Sect. \ref{sec: GRMHD}, we introduce the underlying GRMHD equations and simulation setups. We describe the tests we conducted on the resolution of the magneto-rotational instability (MRI). Thereafter, we explain our study of the impact of the initial magnetic energy in the torus on the jet energetics and on the efficiency of energy extraction from the BH and disk to the energy flux in the jet. We connect this analysis further with the investigation of the angular momentum transport of the five simulations. In connecting it to the morphological study of the radiative propertities of a subset of three simulations in Sect. \ref{sec: GRRT}, we considered the impact of magnetic-flux eruptions from the BH on the hydrodynamical jet structure of these simulations. In Sect. \ref{sec: GRRT}, we describe how we ray-traced the polarized synchrotron emission from a population of thermal and nonthermal electrons asserted by the GRMHD simulations. After the ray-tracing setup is introduced, we compare the jet morphology at 86 GHz and present spatially resolved, as well as time-dependent, linear polarization fractions. Finally, we compare the jet edge and width profiles of the three different models with observations from \cite{Lu_2023} and discuss the impact of flux eruptions on the synchrotron emission from the jet and disk.

\section{General-relativistic magnetohydrodynamics}\label{sec: GRMHD}

We employed the 3D GRMHD code \textit{\emph{Kokkos-based High-Accuracy Relativistic Magnetohydrodynamics with AMR}} (\texttt{KHARMA}; \cite{Prather2021, prather2024}), which simulates accretion onto and jet launching from BHs.
\texttt{KHARMA} solves the ideal magneto-hydrodynamics (MHD) equations on curved space times with a four-metric $g_{\mu\nu}$ and a metric-determining $g$ in geometric units ($G=c=1,$ and a factor of $\nicefrac{1}{\sqrt{4\pi}}$ is absorbed in the magnetic field). Throughout this paper, we use code units for the time and length scales; i.e., the light-crossing time is redefined as $t_\mathrm{g} = 1\, G_\mathrm{N} M_\mathrm{BH}/(c^3) \equiv 1\, \unit{M,}$ and the gravitational radius is $r_\mathrm{g} = 1\, G_\mathrm{N} M_\mathrm{BH}/(c^2) \equiv 1\, \unit{M}$, where $G_\mathrm{N}$ is Newton's constant, $c$ is the vacuum speed of light, $M_\mathrm{BH}$ is the BH mass, and $\unit{M}$ is the mass unit that scales the units in the simulations. Greek indices run through $(0,1,2,3)$ and Roman indices through $(1,2,3),$ and we made use of the Einstein summation convention. We used a Kerr black-hole solution on a modified spherical coordinate system including funky-modified--Kerr--Schild coordinates (FMKS). These coordinates exhibit a decreasing grid cell size in the $\theta$-direction toward the equator ($\theta=\nicefrac{\pi}{2}$) that becomes increasingly cylindrical along the jet-axis ($\theta=0$ and $\theta=\pi$). The radial extent of the cells increases exponentially; therefore, high resolution is given at small radii, where finely resolving the accretion process is vital. ($r$, $\theta$, $\phi$) refer to the ordinary spherical coordinates. The simulation grids are equal in all simulations, in which we employed $N_r = 256$, $N_\phi=128,$ and $N_\theta=128$ cells in the respective directions. The radial boundaries range from $r = (1.1659-1000)\,\unit{M}$ and otherwise use the complete map in $\phi=(0-2\pi)\,\unit{rad}$ and $\theta=(0-\pi)\,\unit{rad}$.
The underlying equations solved by \texttt{KHARMA} are the covariant conservation of mass, local conservation of energy-momentum, and the covariant Maxwell equations (see \cite{mizuno2024} for more details on the derivation):

\begin{equation}
    \nabla_\mu(\rho u^\mu)=0; \hspace{15pt} \nabla_\mu T^{\mu\nu}=0; \hspace{15pt} \nabla_\mu\tensor[^*]{F}{^\mu^\nu}.
\end{equation}
$\rho$ denotes the rest-mass density and $u^\mu$ the four-velocity of the fluid. $\tensor[^*]{F}{_\mu_\nu}$ is the dual Faraday tensor, and the stress-energy tensor is defined in the case of ideal MHD as follows:
\begin{equation}\label{eq: stress-energy tensor}
    T^{\mu\nu}=(\rho h_{\mathrm{tot}})u^\mu u^\nu + \left( p+\frac{1}{2}b^2 \right)g^{\mu\nu} - b^\mu b^\nu,
\end{equation}
with a total specific enthalpy of $h_{\mathrm{tot}}=h+\nicefrac{b^2}{\rho}$ and a fluid pressure of $p$. The magnetic-field four-vector $b^\mu$ in its temporal component and spatial components reads
\begin{equation}\label{eq: magnet. four-vec.}
    b^t=\frac{\Gamma (B^i u^\mu)}{\alpha}; \hspace{15pt} b^i= \frac{(B^i+\alpha b^t u^i)}{\Gamma},
\end{equation}
where $B^i$ are the spatial components of the magnetic-field three-vector measured by an Eulerian observer, $\alpha$ is the lapse function, and $\Gamma$ is the bulk Lorentz factor. Note that with the definition of $b^\mu$ in Eq. \eqref{eq: magnet. four-vec.}, we made use of the notation $b^2=b^\mu b_\mu$ in Eq. \eqref{eq: stress-energy tensor}. We used the plasma beta, $\beta= \nicefrac{2p}{b^2}$, and magnetization, $\sigma=\nicefrac{b^2}{\rho}$, parameters for diagnostics with
\begin{equation}
    b^2=\frac{B^2+\alpha^2(b^t)^2}{\Gamma^2}=\frac{B^2}{\Gamma^2}+(B^i v_i)^2
    .\end{equation}
The system of equations is closed by an equation of state. We assumed an ideal gas that relates the enthalpy, $h,$ to the pressure, $p,$ and density, $\rho$:
\begin{equation}
    h=1+\frac{\hat{\gamma}p}{(\hat{\gamma}-1)\rho},
\end{equation}
where the adiabatic index is $\hat{\gamma}$ (\cite{Anton_2006}).\\
We employed inflow and outflow boundary conditions in a radial coordinate. Along the polar boundaries, we used a solid reflective wall (\cite{2019_Olivares}). In the azimuthal direction, we employed periodic boundary conditions to all physical quantities across the cells at $\phi=0$.
The GRMHD equations were solved with finite-volume and high-resolution shock-capturing methods including the Lax-van Leer Friedrichs (LLF) flux formula in combination with the WENO5 reconstruction scheme and in synchronization with the IMEX scheme. The density floor was set to $\rho_\mathrm{min}=10^{-6}$ and the internal energy floor to $u_\mathrm{min,geo}=10^{-8}$; the ceiling of the magnetization was $\sigma_\mathrm{max}=100,$ and the ceiling of the internal energy was $\left(\nicefrac{u}{\rho}\right)_\mathrm{max}=2$. The initial pressure maximum in the torus was set to $\rho_{\rm{max}}=1$.
The 3D GRMHD simulations were initialized with a magnetized torus in hydrodynamic equilibrium with a rotating Kerr BH in the center of the simulation domain. Here, we used a setup that allows the disk to become partially magnetically arrested. In the MAD setup, we added a single poloidal loop in the vector potential on top of the FM torus (\cite{1976FishboneMoncrief}) in the following way:
\begin{equation}\label{eq: initial vector potential}
    A_\phi\propto max(q - (1-F),0),
\end{equation}
with
\begin{equation}\label{eq: density distribution}
    q=\frac{\rho}{\rho_\mathrm{max}}\left( \frac{r}{r_\mathrm{in}} \right)^3 \sin^3\hspace{-2pt}\theta\, \exp{\left(\frac{-r}{400}\right)}.\end{equation}

\begin{figure}
    \centering
    \includegraphics[width=1.\linewidth]{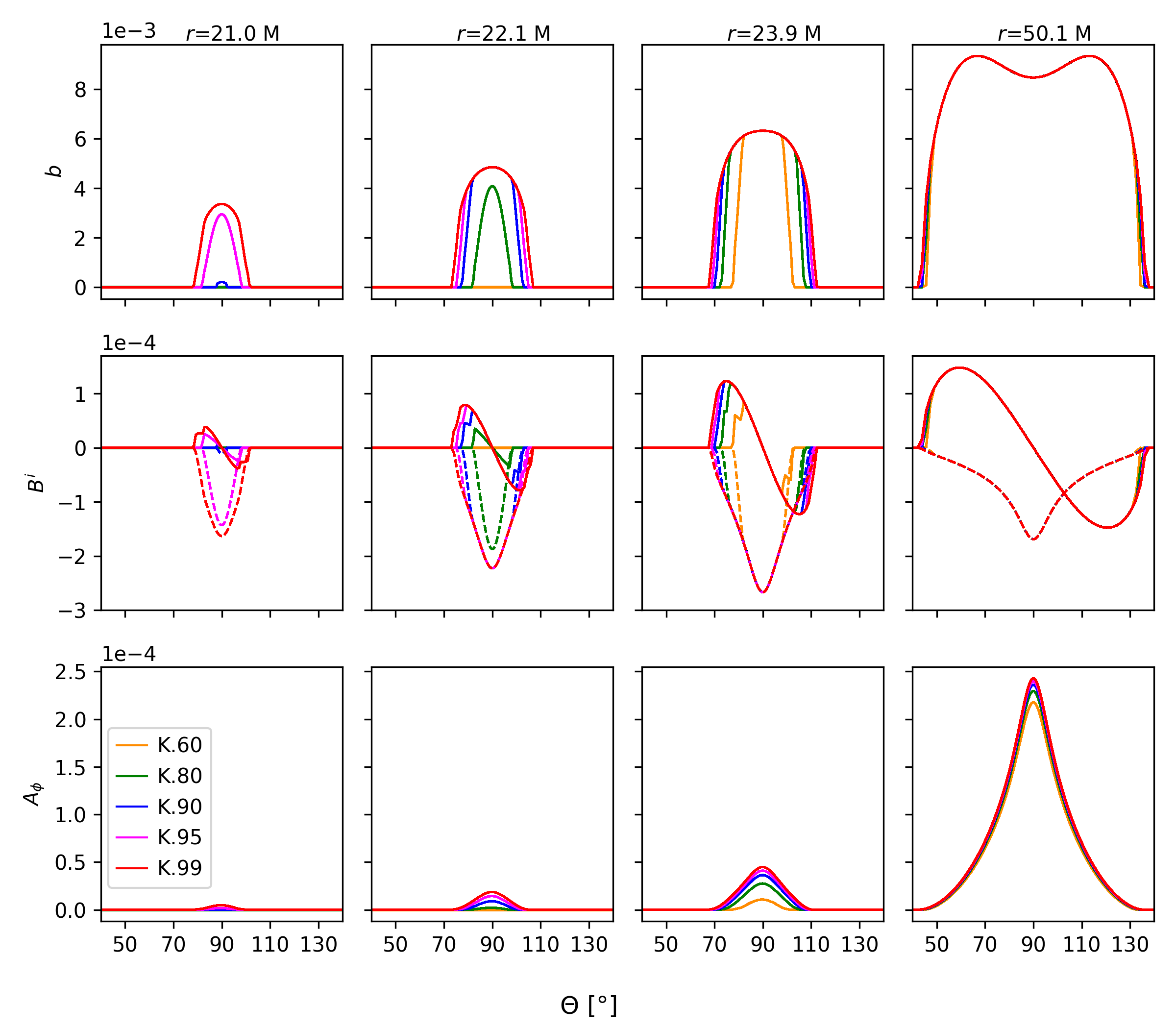}
    \caption{Invariant magnetic-field strength, $b$; the spatial magnetic-field components, $B^r$ and $B^\theta$; and four-vector-potential component, $A_\phi,$ over the polar angle at the initialization step ($t=0\,\rm{M}$). Colors correspond to simulations \texttt{K.60} (orange), \texttt{K.80} (green), \texttt{K.90} (blue), \texttt{K.95} (magenta), and \texttt{K.99} (red). In the middle row, the solid lines show $B^r$ and the dashed lines $B^\theta$. We chose polar slices at radii close to the inner edge of the torus at $r_{\rm{in}}=20\,\rm{M}$ and at the pressure maximum in the tori at $r=50\,\rm{M}$. Note that in columns going from left to right, i.e., radially outwards, the simulations with higher filling factors show increased magnetic-field strengths and extended magnetic field, both radially and in the polar-angle, especially at the inner edge of the torus.}\label{fig: initial A+B-field}
\end{figure}

\noindent
We solely varied the filling factor, $F,$ throughout the simulations from $0.6-0.99$, keeping other initial conditions of the GRMHD simulations constant. Those are a constant adiabatic index, $\hat{\gamma}=\frac{5}{3}$; inner radius, $r_\mathrm{in} = 20;$ and radius of the density maximum, $r_\mathrm{c} = 41,$ of the torus as well as the dimensionless spin parameter, $a=15/16,$ of the BH. The plasma beta was initialized such that its minimal value was above $\beta_{\rm{min}}=2\,\nicefrac{p}{b^2}=100$. We refer to the five models as \texttt{K.60}, \texttt{K.80}, \texttt{K.90}, \texttt{K.95,} and \texttt{K.99;} these correspond to the simulations in that the filling factor is set to $F=0.60$, $F=0.80$, $F=0.90$, $F=0.95,$ and $F=0.99$, respectively. In Appendix \ref{sec: Appendix Initial B-field}, we present more details on the correspondance of the filling factor to the volumes initialized with the magnetic field relative to the torus volume. \\
In Eq. \eqref{eq: initial vector potential}, the effect of $F$ is to proportionally scale the offset from the zero cutoff of $A_{\phi}$, which is visible in the bottom row of Fig. \ref{fig: initial A+B-field}. With increasing $F$, the offset of $A_\phi$ increases, which in turn increases the poloidal extent or, to be precise, the total volume containing cells that have nonzero magnetic-field values. For further details, we refer the reader to Appendix \ref{sec: Appendix Initial B-field}, where we also address that the filling factor does not scale the poloidal-field extent linearly to the extent of the torus; it is rather close to the latter, and for $F\gtrsim0.80$ it is above it (see Tab. \ref{tab: actual torus fillings}).

\begin{figure}[!h]
    \centering
    \includegraphics[width=\linewidth]{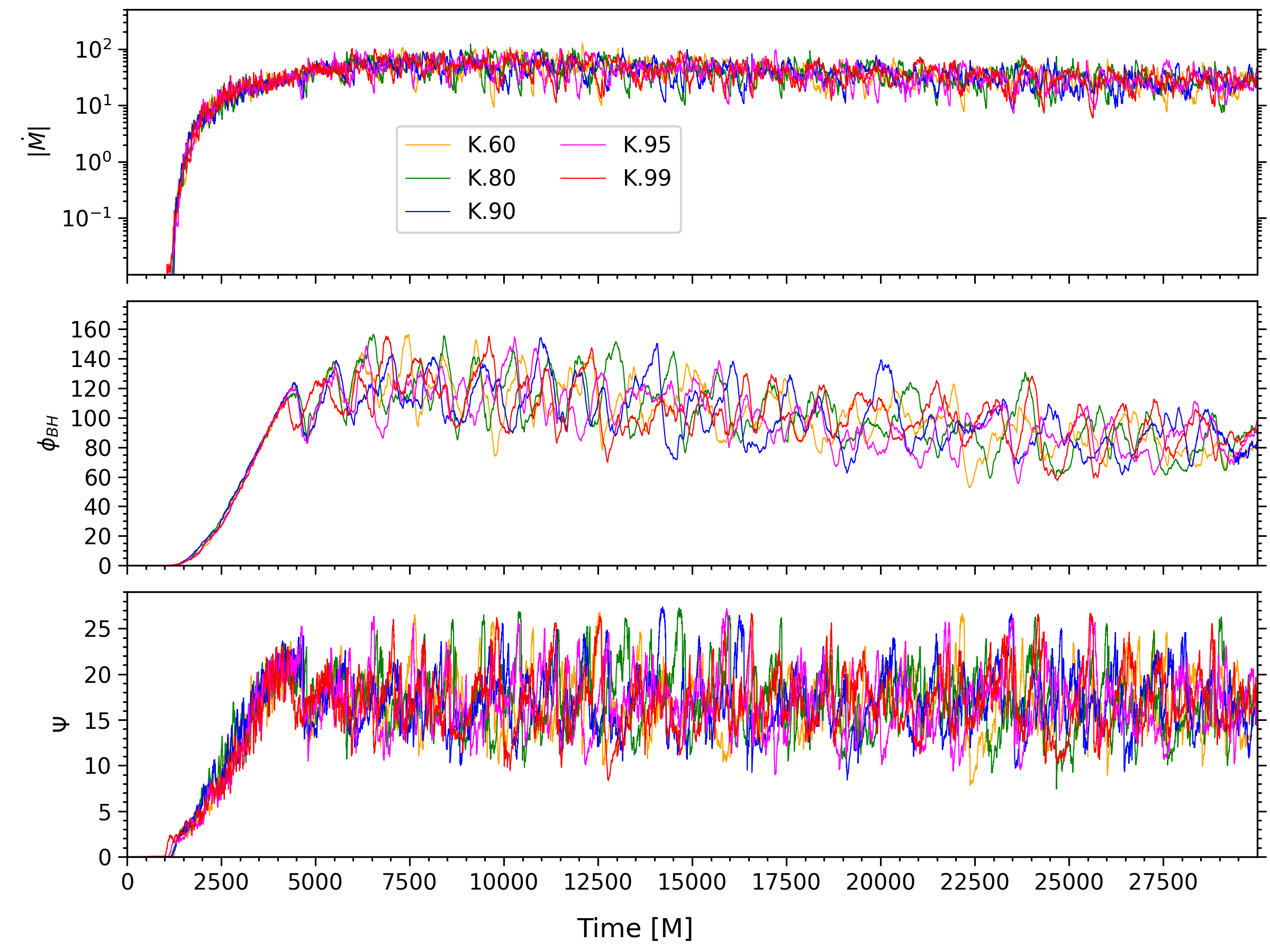}
    \caption{Mass-accretion rates, $\dot{M}$, and magnetic flux through the event horizon of the BH, $\phi_\mathrm{BH,}$ and the MAD-parameter, $\Psi=\phi_\mathrm{BH}/\sqrt{\dot{M,}}$ for all simulations over time.}
    \label{fig: mdot phi MAD-parameter}
\end{figure}

\noindent All simulations were calculated up to a simulation time of $30000\,\unit{M}$. We show the mass accretion rate, $\dot{M}$; the magnetic flux through the event horizon, $\phi_\mathrm{BH}$; and the MAD parameter, $\Psi=\nicefrac{\phi_\mathrm{BH}}{\sqrt{\dot{M}}}$, over the full temporal range of the simulations in Fig. \ref{fig: mdot phi MAD-parameter}.
In Appendix \ref{sec: Appendix MHD parameters}, Fig. \ref{fig: time and azimuthal average plasma K.99}, we present azimuthal and time-averaged cross-sections of the meridional plane of $\rho$, $\sigma$, magnetic field magnitude $b=\sqrt{|b_\mu b^\mu|}$, plasma-$\beta,$ and the ion temperature, $\Theta_i$.

\subsection{Resolution and MRI suppression tests}

In this section, we describe our investigation of the MRI resolution, which is thought to be the main driver of angular momentum and energy transport of plasma accretion in SMBH systems. Due to the limitations of finite resolution, the spectrum of MRI modes is truncated in simulations; hence, it is able to produce various numerical artifacts. Truncation of the MRI modes affects the wave-number range of turbulence modes caused by the MRI and consequently alters the rate at which nonlinear mode--mode couplings transfer energy from large- to small-scale dynamics. Insufficient resolution of the fastest growing MRI mode brings the turbulent field amplification to a halt, and numerical dissipation takes over; hence, the field decays. We tested wether underlying MRI modes were well resolved by the physical criterion and that the number of grid cells that are able to capture the MRI wavelength $\lambda_\mathrm{MRI}$ was sufficiently large.  The so-called quality or $Q$ factors  (\cite{Sano_2004,Noble_2010,Porth_2019}) were defined to reflect the resolution of the underlying MRI mode with the grid spacing in a given spatial direction, $\Delta x^{(i)}$:
\begin{equation}
    Q^{(i)} \equiv \frac{\lambda^{(i)}_\mathrm{MRI}}{\Delta x^{(i)}}.
\end{equation}
The wavelength of the fastest growing MRI mode and the cell extent in a given direction can be defined as
\begin{equation}
    \lambda^{(i)}_\mathrm{MRI} \equiv \frac{2\pi}{\sqrt{(\rho h + b^2)\Omega}} b^\mu e_\mu^{(i)}, \hspace{15pt} \Delta x^{(i)} \equiv (\Delta x^\mu)^{(i)} e_\mu^{(i)}.
\end{equation}
$\Omega=u^\phi/u^t$ is the azimuthal angular velocity of the fluid measured in the locally non-rotating reference frame (LNRF) in Kerr space time, and $e_\mu^{(i)}$ is the $i$-th one-form of the co-vielbein, determining the tetrad basis, where the LNRF is boosted into the comoving frame of the fluid (see \cite{Takahashi_2007} for the basis transformations). $\Delta x^{(i)}$ is the grid spacing in a fluid frame tetrad basis oriented along the LNRF. $(\Delta x^\mu)^{(i)}$ is a vector containing, for each $i,$ one of the grid spacings in the Kerr--Schild coordinates. For the $\theta$ coordinate and grid spacing, $\Delta \theta,$ in Kerr--Schild this would look as follows: $(\Delta x^\mu)^{(\theta)} = (0,0,\Delta \theta, 0)^T$. This is analogously calculated for the other coordinates $(r,\phi)$. We calculated the quality factors $Q^{\theta}$ and $Q^{\phi}$, and we averaged them over a region that includes the wind and disk regions, which is defined by the bounds $\pi/3 \geq \theta_{\mathrm{disk}} \geq 2\pi/3$ and the radial upper bound $r_\mathrm{disk} \leq 50 \,\unit{M}$, where we always took the full azimuthal map of $0 \geq \phi > 2\pi$. The spatial average is calculated as the expectation value over the covariant volume as follows:
\begin{equation}
    \langle Q^{(i)} \rangle_w = \frac{\int\hspace{-5pt}\int\hspace{-5pt}\int \sqrt{-g} \mathrm{d}r\mathrm{d}\theta\mathrm{d}\phi\, Q^{(i)}(r,\theta,\phi) \, w(r,\theta,\phi)}{\int\hspace{-5pt}\int\hspace{-5pt}\int \sqrt{-g} \mathrm{d}r\mathrm{d}\theta\mathrm{d}\phi\, w(r,\theta,\phi)},
\end{equation}
\noindent
where $w(r,\theta,\phi)$ is a weighting function that acts as a masking function. In the case of the disk averages, this function is 1 inside the poloidal bounds specified above and 0 outside of them. \\
We show the temporal evolution of the disk's averaged quality-factors and their median values in Appendix \ref{sec: Appendix Q-factors}, Fig. \ref{fig: Q-factors temporal}.
\cite{Sano_2004} suggested $Q$ factors of $Q > 6$ to correspond to a sufficient resolution of the MRI. We matched these requirements in $Q^{(\theta)}$ and $Q^{(\phi)}$ with consistent values in all simulations with median values in time of $Q_\mathrm{med}^{(\theta)} \gtrsim 35, Q_\mathrm{med}^{(\phi)} \gtrsim 12$ and time averages of $\langle Q^{(\theta)} \rangle_t \gtrsim 60, \langle Q^{(\phi)} \rangle_t \gtrsim 16,$ which are valid for all simulations over the time span $[10000,30000]\,\rm M$ (exact values are shown in Fig. \ref{fig: Q-factors temporal}, and meridional snapshots are given in Fig. \ref{fig: Q-factors snapshot and azimuthal average}).

\subsection{Electromagnetic and kinetic energy flux in the jet}\label{sec: jet energy fluxes}

\begin{figure}[!h]
    \centering
    \includegraphics[width=\linewidth]{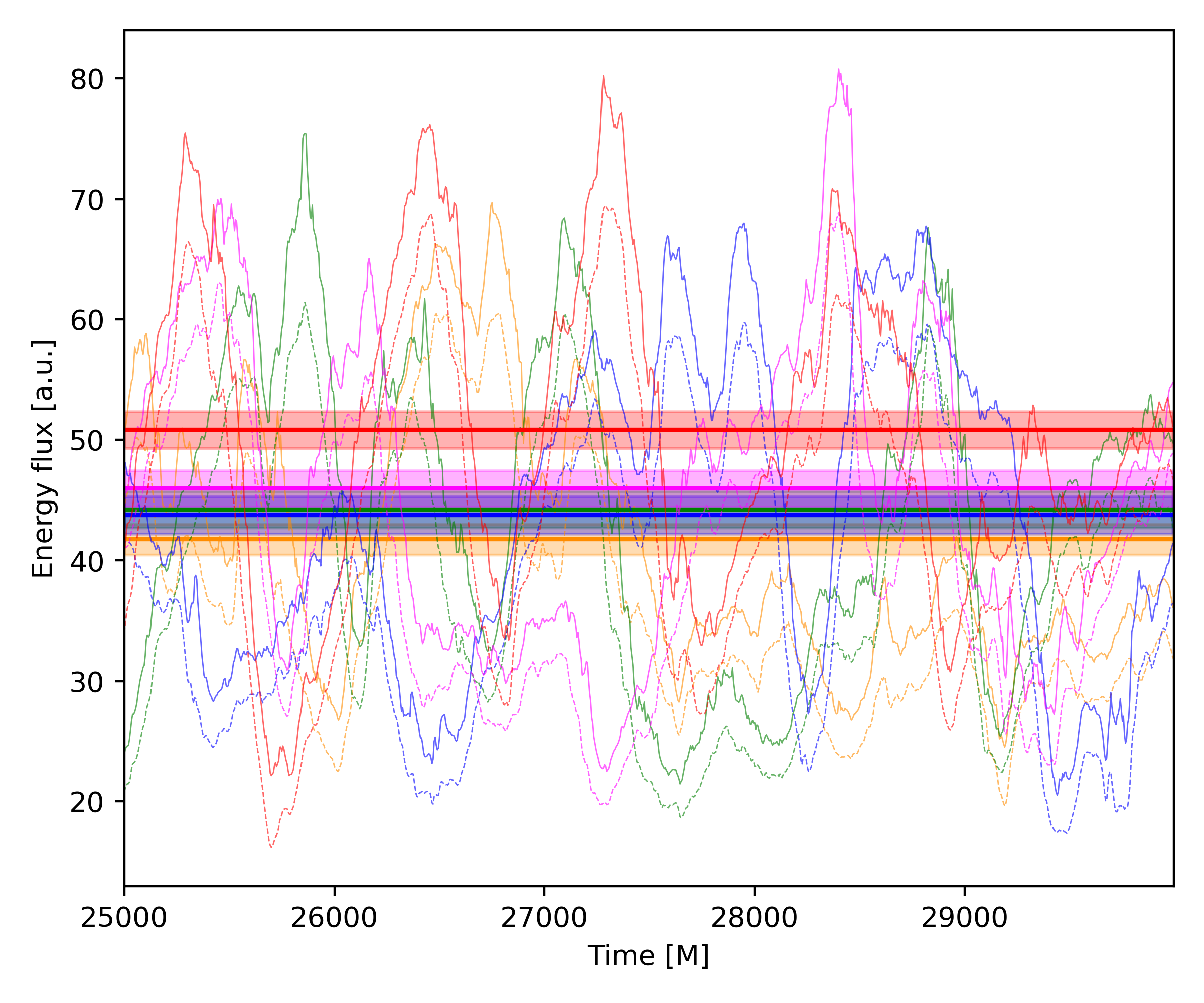}
    \caption{Temporal evolution of total energy flux (solid lines) and EM energy flux (dashed lines) measured at $r=10\, \unit{M}$ in the jet for simulations \texttt{K.60} (orange), \texttt{K.80} (green), \texttt{K.90} (blue), \texttt{K.95} (magenta), and \texttt{K.99} (red). The horizontal lines show the average of the total energy fluxes for each run over the shown time interval.}
    \label{fig: Ltot+EM Jet temporal}
\end{figure}

\begin{figure}[!h]
    \centering
    \includegraphics[width=\linewidth]{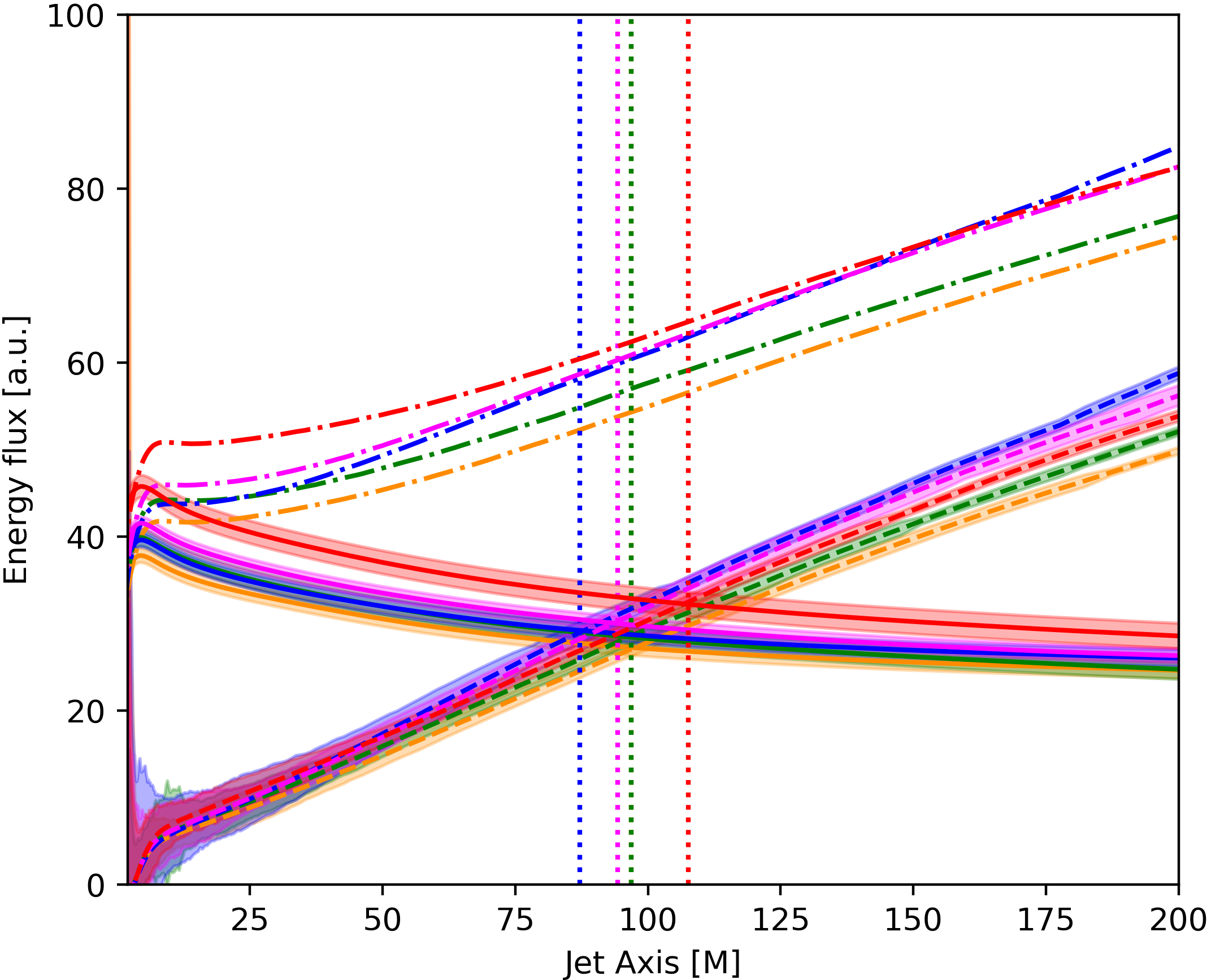}
    \caption{Time-averaged total energy fluxes (dash-dotted lines), EM energy fluxes (solid lines), and kinetic energy fluxes (dashed lines) in the time interval $(25000-30000)\,M$ for varying BH separations along the jet for simulations \texttt{K.60} (orange), \texttt{K.80} (green), \texttt{K.90} (blue), \texttt{K.95} (magenta), and \texttt{K.99} (red). The vertical dotted lines indicate the radii at which the kinetic and EM energy fluxes are in equipartition.}
    \label{fig: Jet-axis L_EM+mass}
\end{figure}

For the time-series analysis presented here, we estimated the error of all time-averaged quantities by means of the correlation error, $\sigma_{\rm{corr}}$ (see Appendix \ref{Appendix Error Estimation} for the formula and more details). This takes into account that the number of data points that are independent of each other is reduced if the data are correlated.
\noindent
The total energy flux is determined via the energy momentum tensor from Eq. \eqref{eq: stress-energy tensor} by contraction with the metric tensor and reformulating the enthalpy in terms of the internal energy density $u=\nicefrac{p}{(\hat{\gamma}-1)}$ to be
\begin{align}
    T^\nu_\mu & = (\rho + \hat{\gamma} u + b^2)\, u^\nu u_\mu \hspace{12pt}- b^\nu b_\mu +(p+\nicefrac{b^2}{2})\delta^\nu_\mu , \\
              & = (\rho + p + u + b^2)\, u^\nu u_\mu - b^\nu b_\mu +(p+\nicefrac{b^2}{2})\delta^\nu_\mu \,,
\end{align}
where $\delta^\nu_\mu$ is the Kronecker delta.
The sub-tensor responsible for the EM contribution to the $r$-$t$ component is
\begin{equation}
    T_{\mathrm{EM}\,t}^{\hspace{11pt}r} =  b^2\, u^r u_t - b^r b_t.
\end{equation}
Analogously, the sub-tensor for the kinetic contributions is
\begin{equation}
    T_{\mathrm{Kin}\,t}^{\hspace{11pt}r} =  (\rho + p + u) \, u^r u_t.
\end{equation}
The total, electromagnetic (EM), and kinetic energy fluxes were then determined by
\begin{align}
    \label{eq: Edot} \dot{E} & =-\int\hspace{-6pt}\int \sqrt{-g} \mathrm{d}\theta\mathrm{d}\phi \, T_t^r                                , \\
    \dot{E}_\mathrm{EM}      & =-\int\hspace{-6pt}\int \sqrt{-g} \mathrm{d}\theta\mathrm{d}\phi \, T_{\mathrm{EM}\,t}^{\hspace{11pt}r}  , \\
    \dot{E}_\mathrm{Kin}     & =-\int\hspace{-6pt}\int \sqrt{-g} \mathrm{d}\theta\mathrm{d}\phi \, T_{\mathrm{Kin}\,t.}^{\hspace{11pt}r}
\end{align}
Before executing the integral, we imposed the condition that the Bernoulli parameter had to be $>1.02$, corresponding to the outflowing part of the plasma shown by the area inside the contours in the bottom row of Fig. \ref{fig: Jet inner and outer edge GRMHD}. We further calculate the energy fluxes at varying radial coordinates and times. \\
In Fig. \ref{fig: Ltot+EM Jet temporal}, we show that the time dependence of the total and EM energy fluxes calculated at a fixed radius of $r=10\,\mathrm{M}$. \texttt{K.99} on average exhibits a significantly higher energy flux of $\sim 50 \,[\mathrm{a.u.}],$ exceeding the other simulations by $\gtrsim 3 \sigma_{\rm{corr}}$; this directly boosts the efficiency of the jet launching due to the comparable mass-accretion rates over all simulations. The energy flux at the jet base is highly dominated by the EM part of the energy flux ($\sim87\, \%$) for \texttt{K.99} and causes the total energy flux excess. The dependence of the time-averaged total energy flux through the jet on the distance from the BH along the jet's parallel axis can be taken from Fig. \ref{fig: Jet-axis L_EM+mass}. The EM and kinetic contributions are in equipartition at different distances from the BH (vertical dashed lines) for each simulation. This is especially evident for simulation \texttt{K.99}, which is significantly more electromagnetically dominated both in extent along the jet and magnitude. We note that the energy fluxes increase with BH separation along the jet, since plasma from the wind region becomes gravitationally unbound; i.e., it reaches $h u_t>1.02$.

\subsection{Outflow efficiencies}\label{sec: jet efficiencies}

\begin{figure}[!h]
    \centering
    \includegraphics[width=\linewidth]{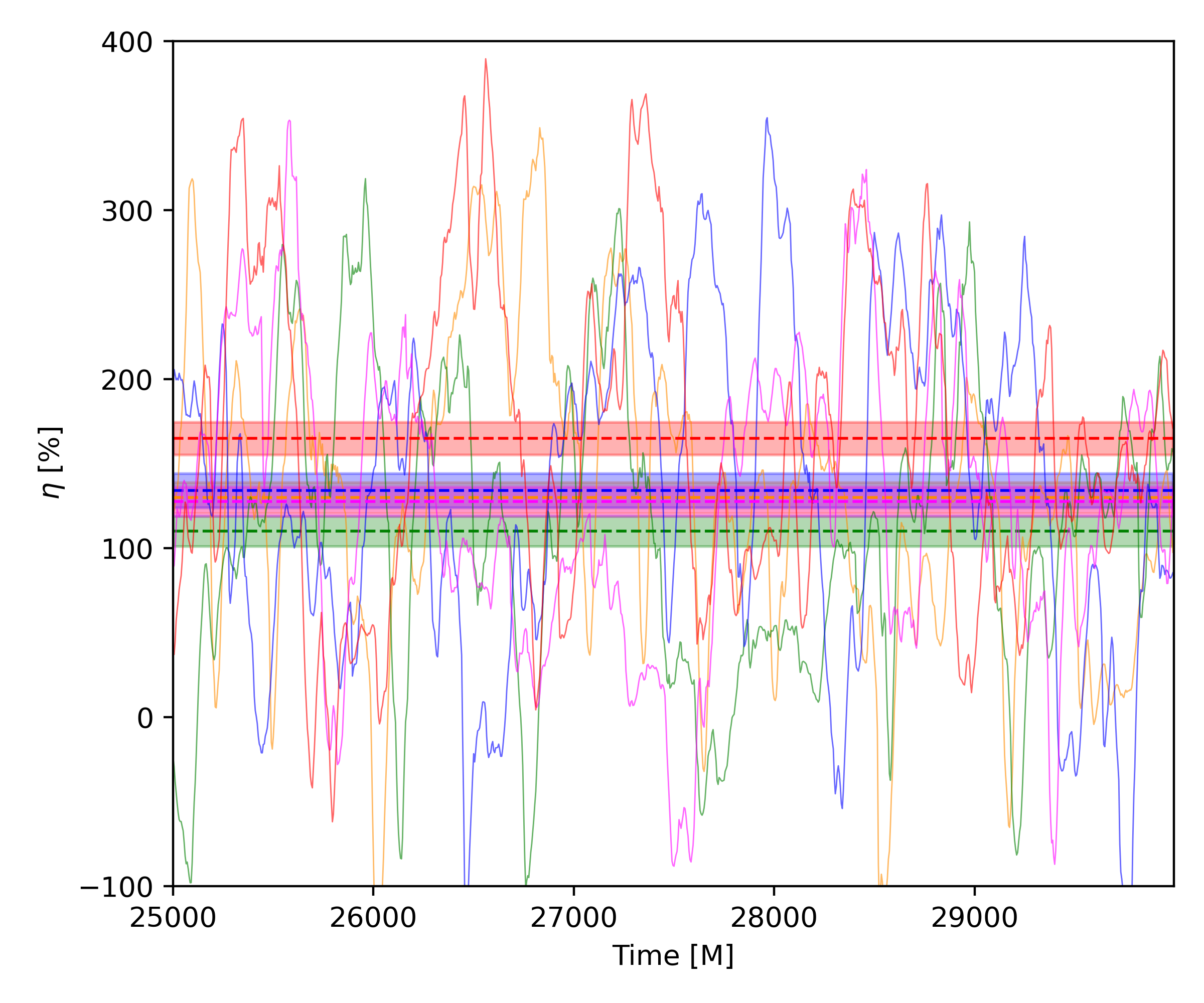}
    \caption{Jet efficiencies for simulations \texttt{K.60} (orange), \texttt{K.80} (green), \texttt{K.90} (blue), \texttt{K.95} (magenta), and \texttt{K.99} (red), where the dashed horizontal lines show the time-averaged efficiencies and the correlation-corrected $2\sigma_{\rm{corr}}$ error band  over the shown interval (listed in Tab. \ref{tab: Outflow Efficiencies of Jet}).}
    \label{fig: Jet efficiency}
\end{figure}

We define the efficiency of energy extraction from the BH spin and disk to outflowing energy at the poles as follows:

\begin{equation}\label{eq: effiency}
    \eta = \frac{\dot{M}_\mathrm{5M} - \dot{E}}{\langle \dot{M}_\mathrm{5M} \rangle},
\end{equation}
where we made use of Eq. \eqref{eq: Edot} and calculated the mass-accretion rate at $r=5\,\mathrm{M}$, $\dot{M}_\mathrm{5M}$ and its averaged $\langle \dot{M}_\mathrm{5M} \rangle$ over the time interval $t=[25000,\, 30000]\,\unit{M}$. This time interval is the most suitable one, as the accretion disk is well converged into a steady state. This can be seen by the mass-accretion rate decelerating its decline toward the end of the simulations (see Fig. \ref{fig: mdot phi MAD-parameter}). For comparison, we also show $\dot{M}_{\rm{5M}}$ in the aforementioned interval in Fig. \ref{fig: Mdot 5M comparison}. The nearly identical average mass-accretion rates also aid the comparison of the efficiencies (see Eq. \eqref{eq: effiency}).

We explicitly did not use the mass-accretion rate at the event horizon, $\dot{M}$ (compare Fig. \ref{fig: mdot phi MAD-parameter}); we instead calculated the flux at $5\,\mathrm{M}$ since this excludes the injected matter to set the density floor at the inner simulation boundary. The radial rest-mass energy density flux is defined as follows:
\begin{equation}
    \dot{M}_\mathrm{5M} = \int\hspace{-6pt}\int_{r=\mathrm{5M}} \sqrt{-g} \mathrm{d}\theta\mathrm{d}\phi \, \rho u^r.
\end{equation}

\noindent
The resulting efficiencies of the energy-extraction process in all three models are shown in Fig. \ref{fig: Jet efficiency}. On average, the \texttt{K.99} simulation exhibits a vastly improved efficiency of $\sim 60\,\%$ excess compared to the commonly used MAD setup, \texttt{K.80}. The averages and peak values of all simulations can be found in Tab. \ref{tab: Outflow Efficiencies of Jet}.%

\begin{table}[!h]
    \sidebysidecaption{0.69\linewidth}{0.3\linewidth}{
        \centering
        \begin{tabular}{|c|c|c|}
            \hline
            Simulation    & $\langle \eta \rangle \, [\%]$ & $\max{(\eta)}\, [\%]$ \\
            \hline
            \texttt{K.60} & 129.8 $\pm$ 9.2                & 348.7                 \\
            \texttt{K.80} & 110.0 $\pm$ 9.0                & 318.6                 \\
            \texttt{K.90} & 134 $\pm$ 10                   & 355                   \\
            \texttt{K.95} & 127.5 $\pm$ 8.7                & 353.2                 \\
            \texttt{K.99} & 164.9 $\pm$ 9.7                & 390.0                 \\
            \hline
        \end{tabular}
    }{\caption{Time averages, corresponding error, and maxima of efficiencies comparing all simulations (cf. Fig. \ref{fig: Jet efficiency}).}\label{tab: Outflow Efficiencies of Jet}}
\end{table}%

\subsection{Angular-momentum transport}\label{sec: angmom transport}

In this section, we investigate the angular-momentum transport by linking it to the jet efficiency and to the impact of the magnetic energy content in the torus on the dynamics from the initialization of the simulations. Following the approach and definitions of the angular-momentum fluxes of \cite{Chatterjee_2022}, we define the radial component of the time-averaged and shell-integrated angular-momentum flux vectors, $\dot{J}^i$ ($i\in(r,\theta)$), accordingly. The total flux is defined as
\begin{equation}
    \dot{J}_{\rm{tot}}^i= \langle T^i_\phi \rangle_{t,\phi},
\end{equation}
and the advective flux reads
\begin{equation}
    \dot{J}_{\rm{adv}}^i= \biggl< \left( \rho+u_{\rm{g}}+b^2/2 \right) u^i \biggl>_{t,\phi} \, \langle u_\phi \rangle_{t,\phi}.
\end{equation}
The stress-induced flux is given by
\begin{equation}
    \dot{J}_{\rm{stress}}^i = \dot{J}_{\rm{tot}}^i - \dot{J}_{\rm{adv}}^i,
\end{equation}
which can be further subdivided into the flux induced by Maxwell-stresses,
\begin{equation}
    \dot{J}_{\rm{stress,M}}^i= \langle b^2\, u^i u_\phi /2 - b^r b_\phi \rangle_{t,\phi},
\end{equation}
and by Reynolds stresses, which are subdominant and therefore do not add anything to the discussions in this paper; hence, $\dot{J}^i_{\rm{stress}} \sim \dot{J}^i_{\rm{stress,M}}$.

\begin{figure*}[!t]
    \centering
    \includegraphics[width=\linewidth]{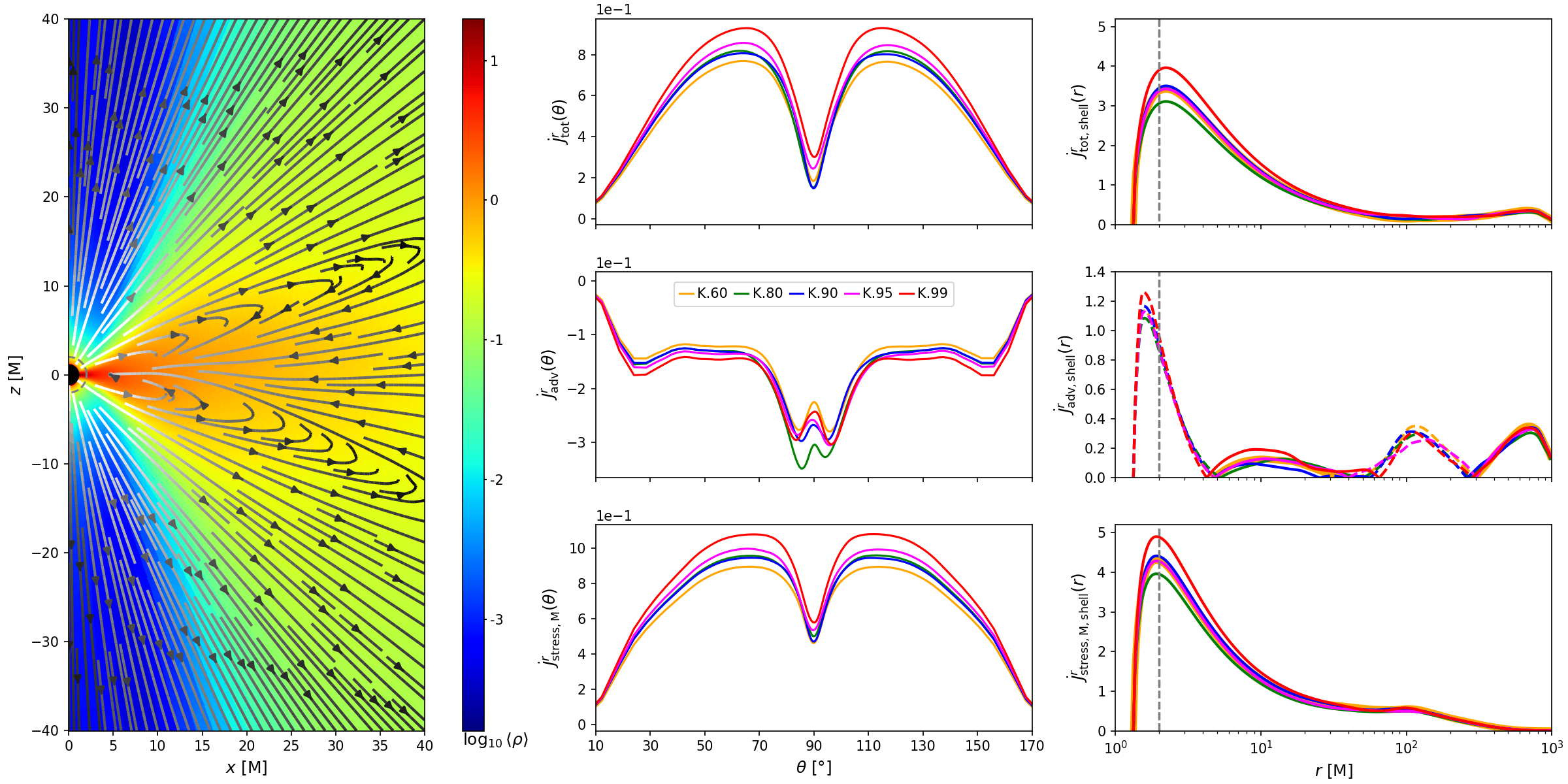}
    \caption{Left panel: Time- and azimuthally averaged density distribution in the meridional plane and total angular-momentum flux streamlines (color-coded by a log-scale of the absolute value; increasing from black to white; simulation: \texttt{K.80}). Middle column: Polar slices of the radial angular-momentum flux at $r=2\,\unit{M}$, indicated by the dashed gray lines in the left panel and right column. Shown are the total, advective and Maxwell-stress-induced angular-momentum fluxes from top to bottom. Right column: Shell-integrated radial component of the angular-momentum fluxes. Solid lines indicate positive values, i.e., outward fluxes and dashed lines are the absolute values of inward (negative) flux.} \label{fig: angmomfluxes}
\end{figure*}%
\noindent
We time-averaged the quantities again over the $t = [25000,\,30000]\,\rm{M}$ interval in order to compare it with the jet energetics and efficiencies from Sect. \ref{sec: jet energy fluxes} and Sect. \ref{sec: jet efficiencies}.\\
\noindent
In Fig. \ref{fig: angmomfluxes}, we present the polar and radial dependence of time- and azimuthally averaged angular-momentum transport. In the left panel, we show the time- and azimuthally averaged total angular-momentum vector field for the first time for simulations in a MAD setup with a high prograde BH spin (cf. \cite[][]{Chatterjee_2022} Fig. 10, showing the case of a non-rotating BH in the MAD-setup). The meridional cross-section of the angular-momentum vector field exhibits quite distinct characteristics from the case of the MAD setup with a non-rotating BH, in that it exhibits a monopole field geometry in the jet and wind region up to a polar angle of $\sim 70^\circ$ and transitioning into a quadrupole field geometry in the disk and torus. Through this magnetospheric configuration, the polar outflow from the jet and wind feeds back into the equatorial inflow in the disk, as opposed to the non-rotating MADs, where there is a transition from an equatorially parallel inflow inside the disk scale height (similar opening angle to the transition to the loops) to the parabolically outflowing angular-momentum flux in the jet. The major difference here is the dominance of the angular momentum transport driven by the rotating BH, where the dominantly radial outflow is redirected toward the equatorial plane into the disk, where the torus has to facilitate the radially outward angular-momentum transport in the wind region.  \\
\noindent
In the middle and right columns of Fig. \ref{fig: angmomfluxes}, we display the impact of the magnetic filling of the MAD torus on the angular momentum transport. The middle column shows the polar dependence of the time- and azimuthally averaged radial component of the angular-momentum vectors at $r=2$ M, where there is a maximum in the shell-integrated fluxes. The right panel shows the shell-integrated (see Sect. \ref{sec: jet energy fluxes} for the definition of shell-integration) and time-averaged radial angular-momentum fluxes. The polar slices (indicated by dashed gray lines in the left panel and right column) of $\dot{J}_{\rm{tot}}^r$ and $\dot{J}_{\rm{stress,M}}^r$ show a clear model trend. Higher initial magnetic energy from the torus leads to an increase in outward angular-momentum flux, which is also sustained at larger radii of $r\lesssim50\,\unit{M}$ (presented in Fig. \ref{fig: polar slices angularmomentum fluxes}).\\
In the right panel at the top, one can see that the total shell-integrated fluxes are positive; i.e., radially outgoing flux in the jet and wind are dominant, whereas for the case of a non-rotating BH in \cite{Chatterjee_2022}, Fig. 14, the shell-integrated radial component of the total angular-momentum flux is constant and negative up to $r\sim100$ M. Secondly, overall the angular-momentum fluxes peak close to the event horizon of the BH, specifically inside the ergosphere, and they drop outside of this region. \\
We note that the model trend is not identical for the shell-integrated total and Maxwell-induced fluxes, since these are affected by the inward advective flux, which is more sensitive to turbulent realizations and does not appear to exhibit any model trend. The model trend is the clearest in the jet sheath as opposed to the equatorial plane in both total and Maxwell-induced angular-momentum fluxes. Looking at the middle panel, one can clearly see that the advective equatorial flux for \texttt{K.80} is the highest; hence, it decreases the total shell-integrated flux the most for this model. However, we observe that simulation \texttt{K.99} overall exhibits the highest angular-momentum flux both inward, driven by advective stresses, and especially outward, where the electromagnetic stresses are dominant. Furthermore, we observe that the model trend in the jet efficiency and the shell-integrated radial angular-momentum fluxes is consistent with one another, which directly links the efficiencies to the angular-momentum transport of these simulations.

\subsection{Flux eruptions}\label{subsec: GRMHD flux eruptions}

\begin{figure}[h]
    \centering
    \includegraphics[width=\linewidth]{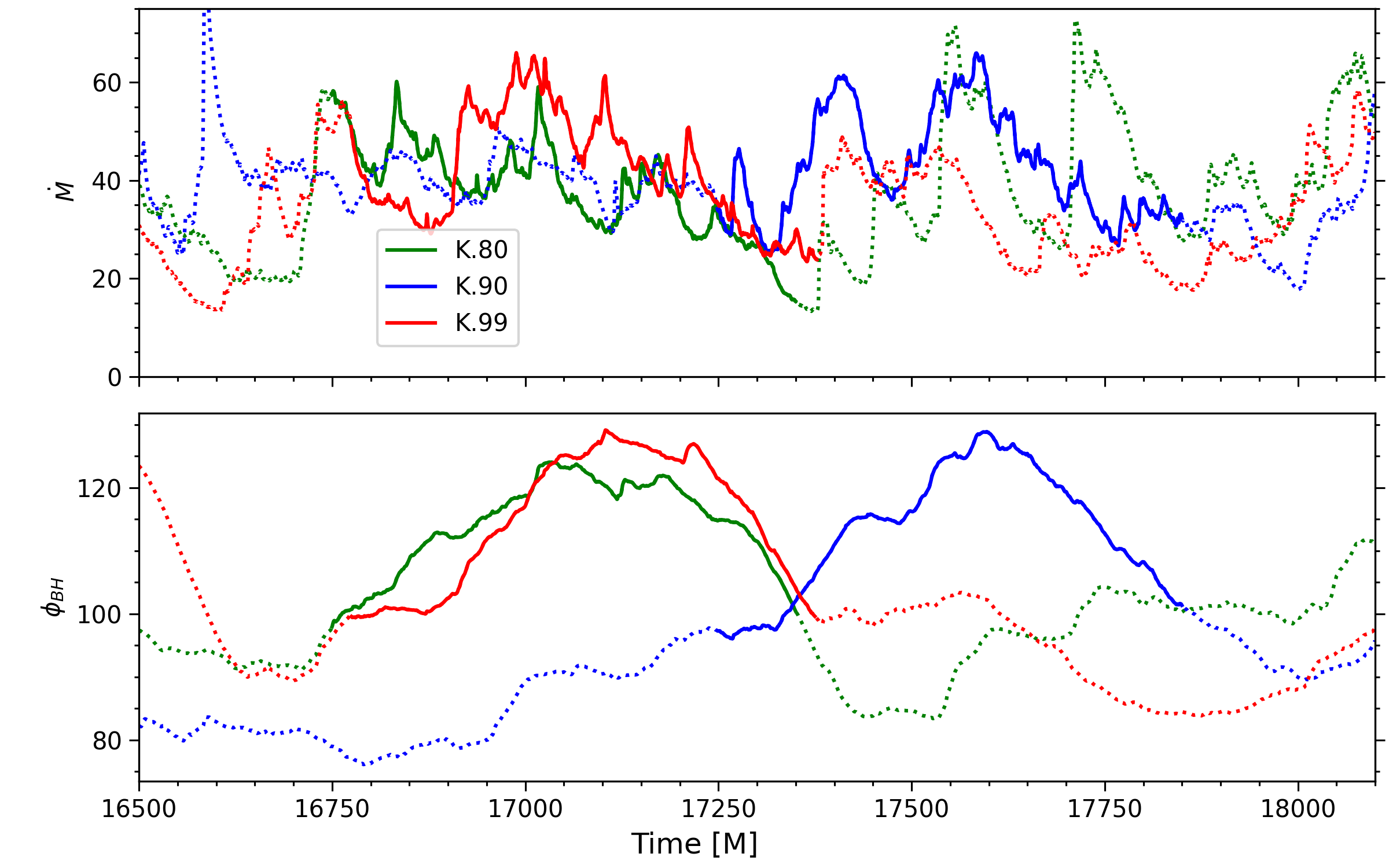}
    \caption{Interval of investigation in order to compare the jet morphologies of all simulations. The magnetic flux, $\phi_\mathrm{BH}$, describes a similar curve in height amplitude and width in time intervals for all simulations (indicated by solid lines). The dashed lines solely indicate time intervals that are not of interest for the discussions of the jet morphology.}
    \label{fig: comparison interval}
\end{figure}

We averaged the data over a time span in which a MAD state occurs in each simulation from a subset of three simulations: \texttt{K.80}, \texttt{K.90,} and \texttt{K.99}. In Fig. \ref{fig: comparison interval}, we use a solid line to emphasize the mass-accretion rate and the magnetic flux over the time spans under investigation for each simulation. The time spans of interest in this study were chosen so that the magnetic fluxes through the event horizon have a similar profile, especially for the maximum and minimum amplitudes to be comparable and the total time span to be $\Delta t = 600\,\unit{M}$. These choices affected the comparability of the different models, especially when interpreting averaged plasma parameters and ray-traced images of azimuthally asymmetric structures, created by the cyclic MAD states. The notion of better comparability in the indicated time intervals in Fig. \ref{fig: comparison interval} is based on the fact that the simulations undergo very similar increases and decreases in accreted magnetic field, which, at the peak of $\phi_\mathrm{BH}$, causes the arrestation of the disk. This arises from an advectively driven phase of accumulation of magnetic field in the vicinity of the BH, which then overreaches the threshold magnetic pressure to be accreted, and outward diffusion of the accreted plasma takes over \citep[][]{jacqueminide2025}. This is observed to occur in each model at very similar evolution states of the simulations (cf. Fig. \ref{fig: comparison interval}). The build up of magnetic flux in the beginning of each interval is very similar, as every simulation accretes in a non-arrested way over a very similar time span, as is given for the duration of the arrested states of the disks in all three simulations. The MAD state (or magnetic-flux eruption) is indicated by the decrease in the mass-accretion rate (cf. Fig. \ref{fig: comparison interval}) after both magnetic fluxes and mass-accretion rates reach their peak in every simulation. We note that these time intervals are the only ones within the whole simulations, where we see such strong similarities in the magnetic flux profiles, especially at those similar and importantly late evolution times ($\sim 17000\;\unit{M}$) in the simulations. Since the electromagnetic contribution of the jet power depends quadratically on $\phi_\mathrm{BH}$ (\cite{2011Tchekhovskoy}), the EM energy flux contributions to the jet can be ruled out from being a result of strongly differing $\phi_\mathrm{BH}$ over the three simulations.

\begin{table}[!h]
    \sidebysidecaption{0.59\linewidth}{0.4\linewidth}{
        \centering
        \begin{tabular}{|c|c|}
            \hline
            Simulation    & Time interval [M] \\
            \hline
            \texttt{K.80} & 16750 - 17350     \\
            \texttt{K.90} & 17250 - 17850     \\
            \texttt{K.99} & 16775 - 17375     \\
            \hline
        \end{tabular}
    }{\caption{Time intervals for the respective simulations that were used for parts of the GRHMD analysis and only used for the GRRT.}\label{tab: Ray-tracing & GRMHD time intervals}}
\end{table}

\begin{figure}[!h]
    \centering
    \includegraphics[width=\linewidth]{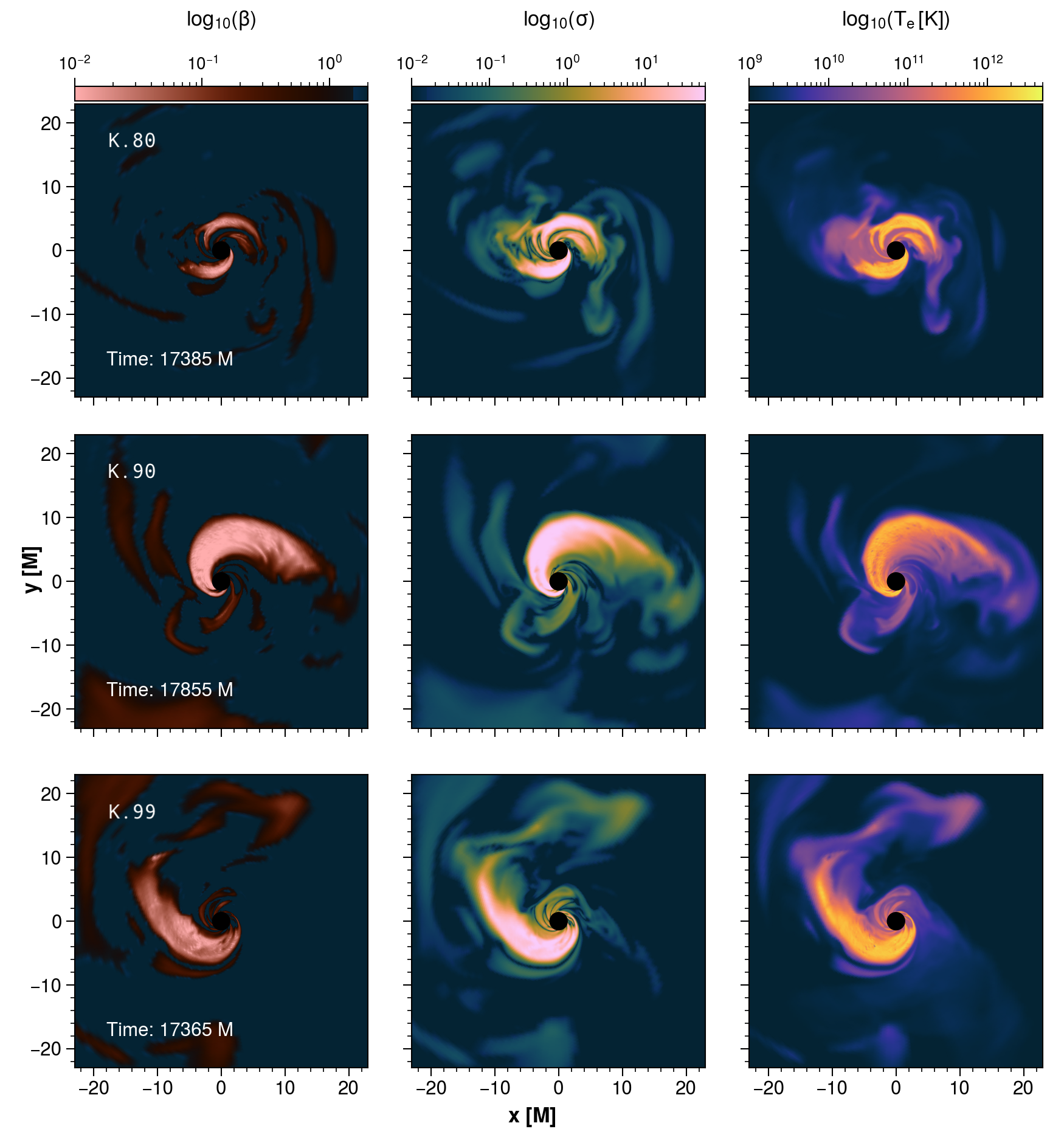}
    \caption{2D slices of the equatorial plane showing plasma $\beta$, $\sigma,$ and $T_\mathrm{e}$ of the peak MAD state of the simulations \texttt{K.80}, \texttt{K.90,} and \texttt{K.99}. Note the extended, arrow-like tail of the flux tube in \texttt{K.99} that is twice the length compared to that of \texttt{K.90}.}
    \label{fig: MAD-state xy}
\end{figure}

\begin{figure}[!h]
    \centering
    \includegraphics[width=\linewidth]{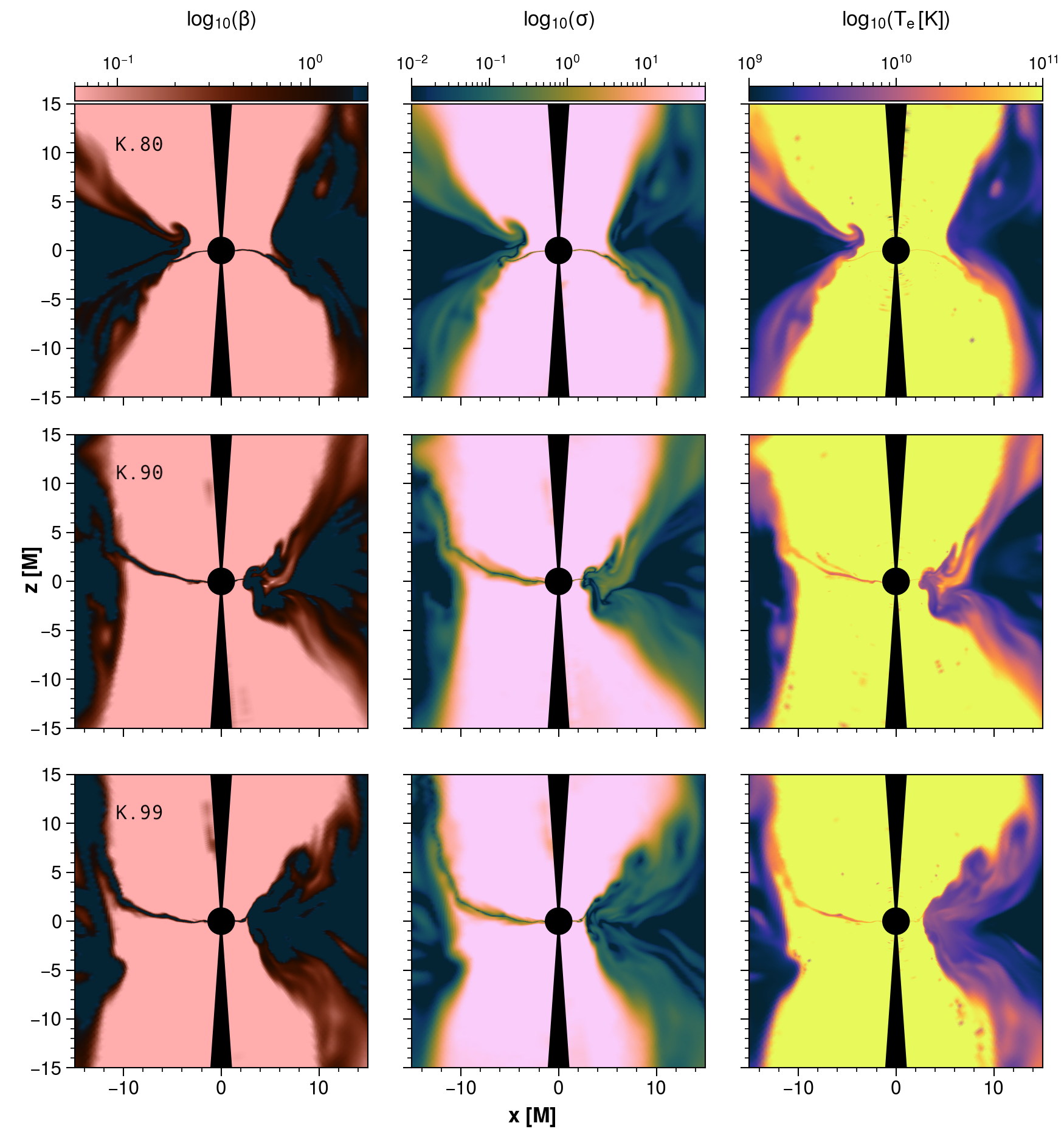}
    \caption{Meridional 2D slices of peak MAD state showing plasma $\beta$, $\sigma,$ and $T_\mathrm{e}$ of the simulations \texttt{K.80}, \texttt{K.90,} and \texttt{K.99} (equal times to those in Fig. \ref{fig: MAD-state xy} for each simulation, respectively). We show the $180^\circ$ azimuthally rotated cross-section of \texttt{K.99}, for comparison of the MAD states.}
    \label{fig: MAD-state xz}
\end{figure}
\noindent
We wanted to characterize the magnetic flux eruptions or MAD states, since their impacts on the morphology are essential for consideration in the subsequent analysis. We show, respectively, simultaneous snapshots of the meridional and equatorial planes in the GRMHD simulations, where the flux eruptions are at their respective peaks in Fig. \ref{fig: MAD-state xy} and Fig. \ref{fig: MAD-state xz}. We note that the comparison of the snapshots lies outside the time intervals of comparison in Tab. \ref{tab: Ray-tracing & GRMHD time intervals}; nevertheless, the MAD states already impact the simulations inside these intervals, and we show the peak MAD states to emphasize these effects. We want to demonstrate the maximal impact, that the flux eruptions can have on the plasma parameters and on the radiative signatures. The timescales from the onset of these magnetically arrested states to their peak flux eruptions and, finally, termination of the arrested disk do not differ significantly across the three models. This in turn aids the comparability of the time intervals (compare Tab. \ref{tab: Ray-tracing & GRMHD time intervals}) chosen for time-dependent and time-averaged quantities. We want to emphasize that although the amplitude of magnetic flux that is built up and released is very similar across all simulations (cf. Fig. \ref{fig: comparison interval}), the flux tubes (cf. \cite{Porth_2019}) do vary strongly in terms of their electron temperature, magnetization, and kinetic-to-magnetic pressure ratio (see Fig. \ref{fig: MAD-state xy} and Fig. \ref{fig: MAD-state xz}). In the simulation \texttt{K.80,} these magnetic eruptions occur two-sided in this instance, whereas both \texttt{K.90} and \texttt{K.99} are completely azimuthally asymmetric (cf. Fig. \ref{fig: MAD-state xy}), and the one-sided flux eruptions are much more stable in the latter simulations. \texttt{K.99} shows a significantly extended flux tube that is hot enough to imprint clear signatures of thermal synchrotron radiation in the ray tracings (see Sect. \ref{sec: GRRT} and Fig. \ref{fig: K.99 Flux tube RT disk and jet}). The jet structure is also highly dependent on the MAD states as these flux tubes are able to contribute to the jet morphology significantly (see Fig. \ref{fig: MAD-state xz}; \texttt{K.99} is again rotated azimuthally by $180^\circ$ to compare the MAD states), and, especially, they introduce high degrees of azimuthal asymmetries, which is important to consider when we investigate time-averaged quantities and morphologies in the following sections.

\subsection{Comparison of the jet structure}\label{sec: GRMHD jet profiles}

\begin{figure}[!h]
    \centering
    \includegraphics[width=\linewidth]{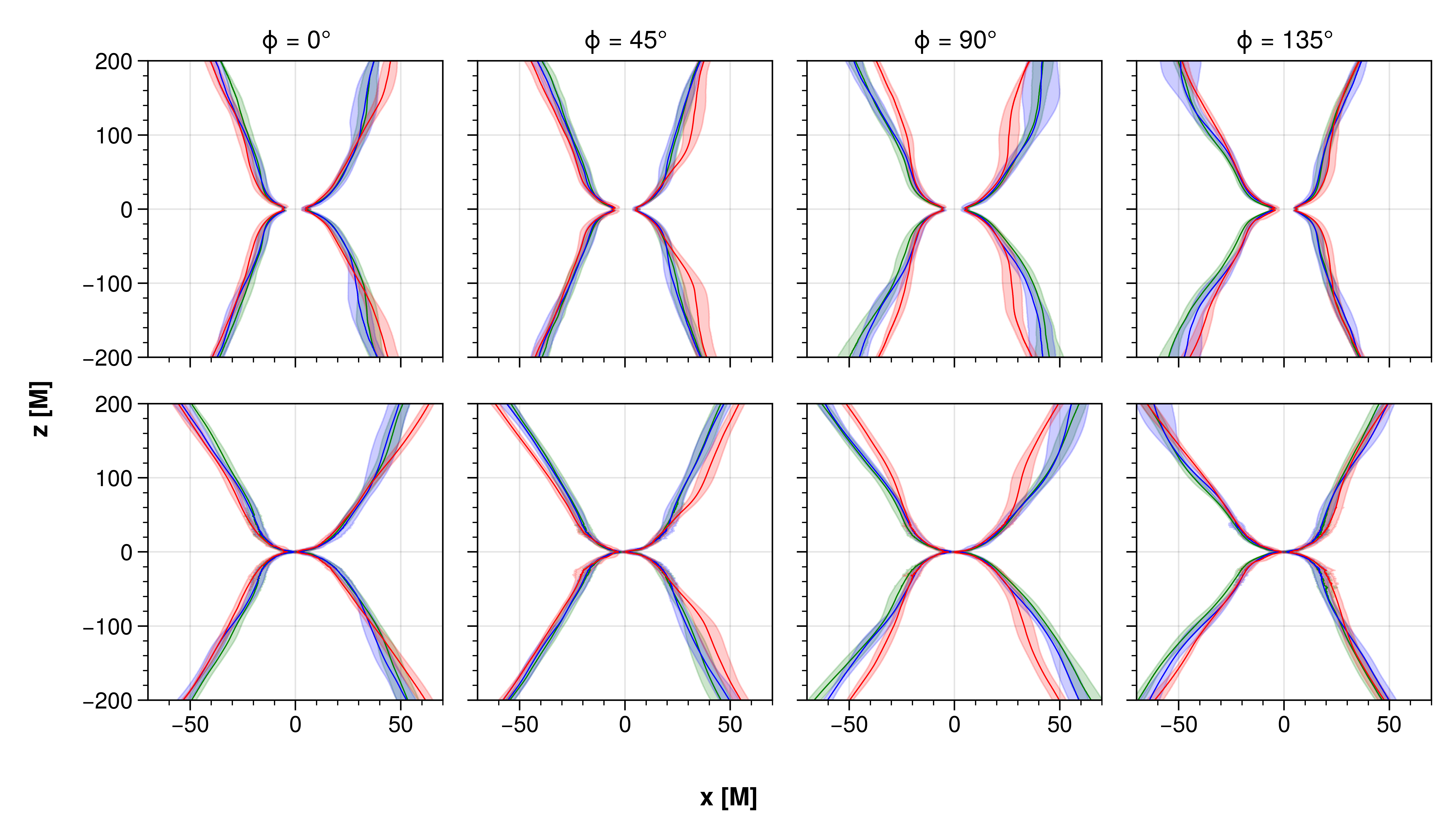}
    \caption{Upper row: $\sigma = 1$ contours in $xz$-cross-section corresponding to \texttt{K.80} (green lines), \texttt{K.90} (blue lines), and \texttt{K.99} (red lines), and the $1\sigma$ standard deviation in the $x$-directions over the time intervals listed in Tab. \ref{tab: Ray-tracing & GRMHD time intervals}. Each column presents poloidal cross-sections at azimuthal angles in $45^\circ$ increments. Bottom row: Bernoulli parameter $-hu_t=1.02$ contours, where we otherwise show the exact analogs for it as in the $\sigma=1$ contours. }
    \label{fig: Jet inner and outer edge GRMHD}
\end{figure}
\noindent
We investigated the jet structure and its variability by means of the magnetization, $\sigma,$ and the Bernoulli parameter, $-hu_t$. The contour at $\sigma = 1 \equiv \sigma_{\mathrm{cut}}$ defines our cutoff of the jet spine in the ray tracings shown in Sect. \ref{sec: GRRT} and can be interpreted as our inner boundary of the jet sheath. The Bernoulli parameter at $-hu_t > 1.02$ separates the outflowing plasma in the jet from inflowing or gravitationally bound plasma in the disk and wind, so we defined the $-hu_t = 1.02$ contour as the outer boundary of the jet sheath. In Fig. \ref{fig: Jet inner and outer edge GRMHD}, we show meridional slices ($x$-$z$-plane) in $45^\circ$ increments in the azimuthal angle, $\phi,$ of the $\sigma_{\mathrm{cut}}=1$ contour (top row) and the Bernoulli parameter contour at $-hu_t = 1.02$ (bottom row), which are averaged in the $x$-direction over the respective time intervals listed in Tab. \ref{tab: Ray-tracing & GRMHD time intervals} for each simulation. Moreover, to emphasize the structural variability we show the standard deviations in the $x$-direction, indicated by the colored areas. Note that by $x$-directions we mean every axis that is perpendicular to the BH spin vector (in our simulations aligned with the $z$-axis). We explicitly did not average over phi, since the MAD states that are undergone in the chosen time intervals (see Fig. \ref{fig: comparison interval}) introduce strong azimuthal asymmetries that would otherwise smear out. \\
In Fig. \ref{fig: Jet inner and outer edge GRMHD}, the runs \texttt{K.80} and \texttt{K.90} trace similar jet boundaries, whereas \texttt{K.99} manifests a consistently wider jet for $\phi = \{0^\circ,45^\circ\}$ and a narrower jet for $\phi = \{90^\circ,135^\circ\}$. This hints at a rather elliptical paraboloid jet geometry in the \texttt{K.99} run, likely induced by the stronger MAD state that causes the azimuthal axis asymmetry. We note that we rotated \texttt{K.99} azimuthally by $180^\circ$, such that the flux eruptions of the simulations match in azimuthal angles. Interestingly, at $\phi = 135^\circ$ all runs consistently show the strongest axis asymmetries, which coincides roughly with the foot point of the flux eruption (compare to \ref{fig: MAD-state xy}, where the x-axis is at $\phi = 0^\circ/180^\circ$).\\
The $1\sigma$ band in Fig. \ref{fig: Jet inner and outer edge GRMHD} shows consistent structural properties as the Bernoulli contour. The sigma contour is overall significantly narrower downstream of the jet and less variable for \texttt{K.99}, especially at distances of $\sim 100\mathrm{M} - 200\mathrm{M}$ from the center, compared to \texttt{K.80} and \texttt{K.90}.

\section{General-relativistic radiative transfer}\label{sec: GRRT}

In order to investigate the radiative signatures of the 3D GRMHD simulations, we calculated the polarized synchrotron emission by means of a general-relativistic radiative-transfer (GRRT) post-processing step. We therefore employed \texttt{IPOLE} (\cite{moscibrodzka2017ipolesemianalyticscheme}): a semi-analytic code for relativistic polarized radiative transfer.

\noindent
We assumed a model for the electron distribution function (eDF), since our GRMHD simulations evolve the electrons neither simultaneously nor separately; rather, they are assumed to tightly follow the evolution of the simulated plasma. The model chosen for this study combines the thermal Maxwell-Jüttner distribution,
\begin{equation}\label{eq: thermal-eDF}
    \frac{1}{n_e}\frac{dn_e}{d\Gamma_\mathrm{e}}=\frac{\Gamma_\mathrm{e}^2\sqrt{1-1/\Gamma_\mathrm{e}^2}}{\Theta_e K_2(1/ \Theta)}\\
    ,\end{equation}
and the nonthermal $\kappa$-eDF (see \cite{Pandya_2016}), which smoothly converges to the Maxwell--Jüttner distribution for low $\Gamma_\mathrm{e}$:
\begin{equation}\label{eq: kappa-eDF}
    \frac{1}{n_e}\frac{dn_e}{d\Gamma_\mathrm{e}}=N\Gamma_\mathrm{e}\sqrt{\Gamma_\mathrm{e}^2-1}\left(1+\frac{\Gamma_\mathrm{e}-1}{\kappa w}\right)^{-(\kappa+1)},
\end{equation}
with the normalization factor $N$; the dimensionless electron temperature $\Theta_e=k_bT_e/m_ec^2$; the modified Bessel function of the second kind, $K_2$; the power-law index $\kappa;$ and the width parameter $w$.\\
The thermal eDF is the relativistic generalization of the thermal Maxwell distribution, and the $\kappa$-eDF was inspired by observations of the solar wind and by results of collisionless plasma simulations (e.g. \cite{2015Kunz}) and converges to the thermal distribution for $\kappa \rightarrow \infty$. Both distributions need an assignment of the electron temperature ,$T_e$, which we define as
\begin{equation}\label{eq: electron temperature 1}
    T_e= \frac{pm_\mathrm{p}}{R_\mathrm{\nicefrac{p}{e}} \,\rho m_\mathrm{e}} = \frac{T_\mathrm{code}}{R_\mathrm{\nicefrac{p}{e}}}\frac{m_\mathrm{p}}{m_\mathrm{e}},
\end{equation}
\noindent
where $k_\mathrm{B}$, $m_\mathrm{p}$, and $m_\mathrm{e}$ are, respectively, the Boltzmann constant and the proton and electron masses. $\rho$ and $p$ are the rest-mass density and pressure from the GRMHD simulation, respectively. The code temperature, $T_\mathrm{code} = \nicefrac{p}{\rho}$, is thus defined by plasma parameters from the GRMHD code as well, and it describes solely the ion temperature in code units. The ratio of proton temperature $T_\mathrm{p}$ and electron temperature $T_\mathrm{e}$, $R$, is defined as
\begin{equation}
    R_\mathrm{\nicefrac{p}{e}}=\frac{T_\mathrm{p}}{T_\mathrm{e}}\equiv R_\mathrm{high}\frac{\beta^2}{\beta^2+1}+R_\mathrm{low}\frac{1}{\beta^2+1},
\end{equation}
with the plasma parameter $\beta=2\frac{p}{b^2}$ provided by the simulation. The so called R-$\beta$ model then has two parameters in total, $R_\mathrm{high}$ and $R_\mathrm{low}$ (see \cite{Mo_cibrodzka_2016}), that together with the code temperature control the electron temperature. Notice, that we have to assert an electron temperature model as the GRMHD simulation does not intrinsically provide an electron temperature.\\
The kappa distribution in Eq. \eqref{eq: kappa-eDF} is defined by its width, $w$, which is now attributed a fraction $\varepsilon$ of the magnetic energy besides the thermal energy (see \cite{Davelaar2019} and \cite{Fromm2022}):
\begin{equation}\label{eq: kappa width}
    w := \frac{\kappa-3}{\kappa}\Theta_\mathrm{e}+\frac{\varepsilon}{2}\left[1+\tanh{(r-r_\mathrm{inj})}\right]\frac{\kappa-3}{6\kappa}\frac{m_\mathrm{p}}{m_\mathrm{e}}\sigma.
\end{equation}

\begin{figure*}[!t]
    \centering
    \includegraphics[width=.75\linewidth]{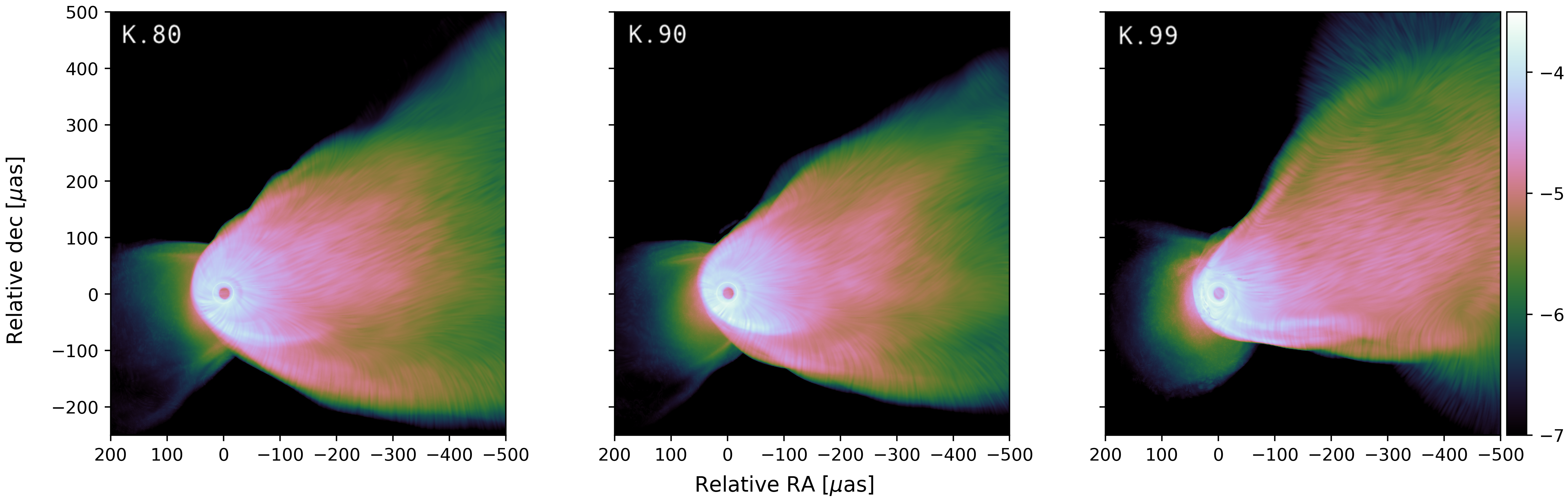}
    \caption{Time-averaged $86$ GHz ray tracings of the M87 jet. Each simulation is averaged over the time intervals corresponding to each simulation that are listed in Tab. \ref{tab: Ray-tracing & GRMHD time intervals}. Stokes I is superimposed with a line integral convolution of the EVPA vector field by scaling the opacity as a function of the linear polarization fraction and rescaling the image by the maximum of the convolved image $\max{[f(I,\, m,\, \mathrm{EVPA}_\mathrm{vector-field})]}$ to $\max{(\mathrm{I})}$.}
    \label{fig: time averaged ray-tracings}
\end{figure*}
\noindent
The injection radius, $r_\mathrm{inj}$, is the distance from the BH, where electrons with magnetic energy contribution are injected and $\sigma=\nicefrac{b^2}{\rho}$. \\ \noindent
The non-thermal contributions to the electron energies are thought to be originated in magnetic reconnection events, in which current sheets accelerate the charged particles. These processes cannot be resolved in global GRMHD simulations, therefore, the $\kappa$-parameter depends on the magnetization and plasma $\beta$ when employing a PIC simulation based sub-grid model from \cite{Ball2018}:
\begin{equation}
    \kappa:=2.8+0.7\sigma^{-\nicefrac{1}{2}}+3.7\sigma^{-0.19}\tanh{\left(23.4\sigma^{0.26}\beta \right)}
\end{equation}
With that we again follow \cite{Fromm2022} and \cite{Davelaar2019} (In the former study, a more detailed parameter space study  on the R-$\beta$ model can be found).\\ \noindent

\subsection{M87 ray-tracing setup}

For the 86 GHz Ray-Tracings, we fixed the parameters of all runs to the mass of the BH of $M_\mathrm{BH}=6.2\times10^9\,\unit{M_\odot}$ \cite[cf.][]{Gebhardt_2011, EHT1, EHTVI, EHTV}, the distance to the BH to $D = 16.9\,\unit{Mpc}$ \cite[cf.][]{Blakeslee_2009, Cantiello_2018, EHTVI, EHTV} and the viewing angle $i = 163^\circ$ (as spanned by the BH spin axis and line of sight) of the BH and jet in M87. The parameters for the electron temperature model are set to $R_\mathrm{high}=160$ and $R_\mathrm{low}=1$ and for the $\kappa$-eDF in Eq. \eqref{eq: kappa-eDF}, we choose a non-thermal emission fraction of $\varepsilon=0.5$ and a injection radius of non-thermal particles at $r_\mathrm{inj}=10\,\unit{M}$. We note that the numerical approximations of the emission and absorption coefficients  are only valid in the range $3.1 \ge \kappa \le 8$, where solely the $\kappa$-eDF is applied, and the Maxwell-Jüttner eDF otherwise.
\\This means that we mainly calculate non-thermal emission in the the jet-sheath and it is noted that the Flux-tubes can also be sufficiently magnetized, such that they emit non-thermally.
\\A polar cut of $\theta_\mathrm{cut}=4^\circ$ (opening angle of the cut-out cone is $\alpha_\mathrm{cone}=8^\circ$) is applied for the ray-tracings as shown before in the GRMHD data (cf. Appendix \ref{sec: Appendix MHD parameters} Fig. \ref{fig: time and azimuthal average plasma K.99} and Fig. \ref{fig: MAD-state xz}). Additionally, we omit the emission from the jet-spine in all regions where $\sigma>\sigma_\mathrm{cut}=1$. This is due to the fact that the highly dominant magnetic energy in the jet-spine is prone to introduce small errors in the total energy evolution, drastically altering the internal energy, hence the plasma temperature.
\\The time-averaged flux of each ray-traced simulation is adjusted such that the flux at horizon scale at 230 GHz matches the observed  $0.9 \,\unit{Jy}$ by the \cite{EHT1} by means of optimizing the mass-accretion rates in the GRMHD simulations.

\begin{figure}[!h]
    \centering
    \includegraphics[width=\linewidth]{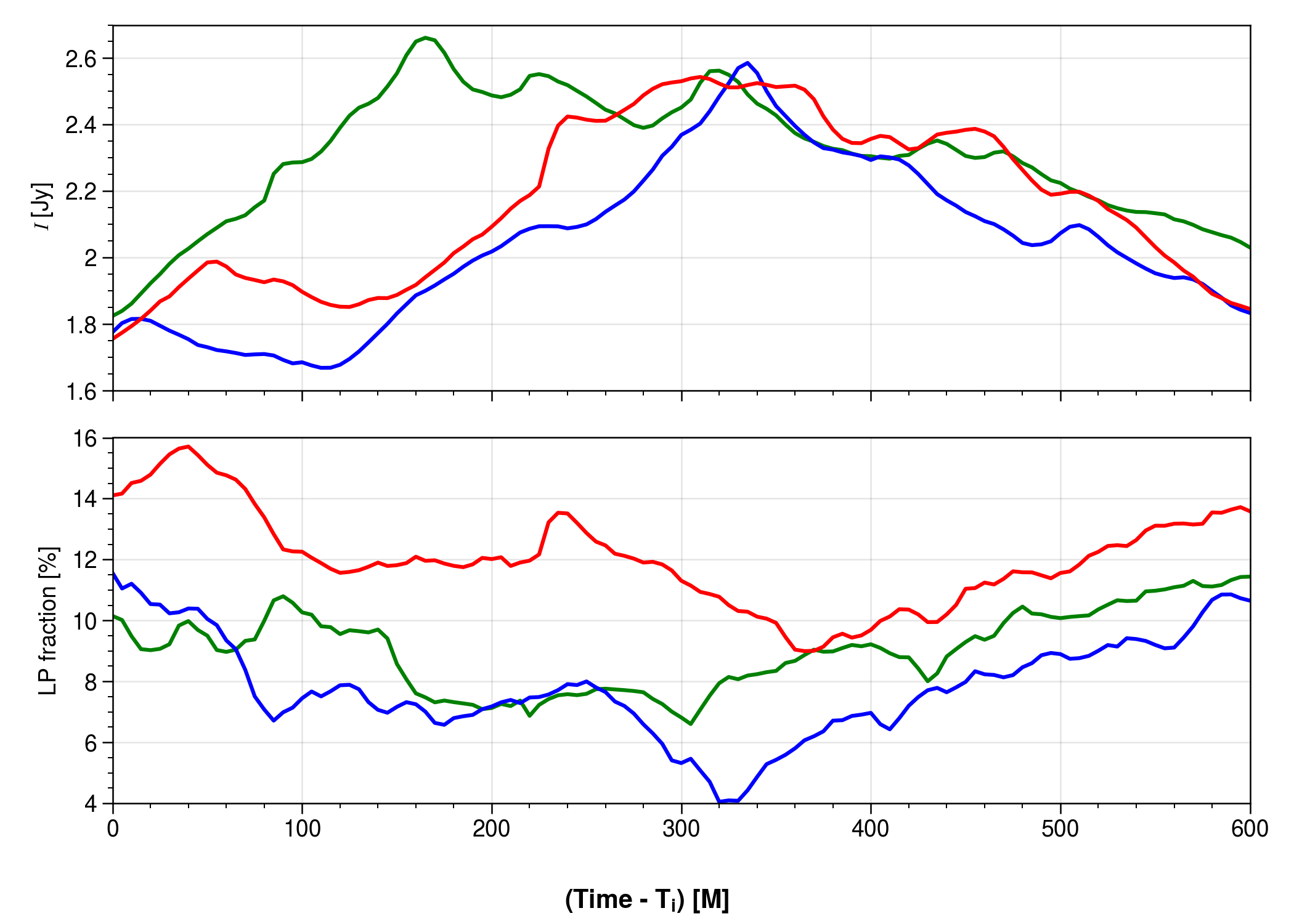}
    \caption{Top row: 86 GHz light curve for simulations \texttt{K.80} (green), \texttt{K.90} (blue), and \texttt{K.99} (red) in the time intervals shown in Tab. \ref{tab: Ray-tracing & GRMHD time intervals}, but shifted by an offset for better comparability. Bottom row: Linear polarization fraction with the same color-code as before.}
    \label{fig: LP fraction (t) and I(t)}
\end{figure}

\subsection{Comparing jet structure and polarization at 86 GHz}

In Fig. \ref{fig: time averaged ray-tracings} we present polarized, time-averaged ray-tracings of the three simulations up to the $0.5\,\unit{mas}$-scale, averaged over the time-intervals as used before (cf. Tab. \ref{tab: Ray-tracing & GRMHD time intervals}). These images are the result of a line-integral convolution (LIC) of the electric vector position angle (EVPA) with a field of random generated numbers. This generates stream lines of the EVPA vector field, encoding directional information of the field lines. The intensity $I$ is then super-imposed onto the LIC of the EVPAs, where the opacity of those is a function of the linear polarization (LP) fraction. The image is then rescaled so that the maximum of the convolved image $\max{[f(I,\, m,\, \mathrm{EVPA}_\mathrm{field-lines})]}$ equals $\max{(I)}$.
The LP fraction is given by:
\begin{equation}
    m=\frac{\sqrt{Q^2+U^2}}{I}\times100 \,[\%]
\end{equation}
\\
Qualitatively, one can tell by the dominance of the stream lines over the smooth Stokes-$I$ in Fig. \ref{fig: time averaged ray-tracings}, that \texttt{K.99} exhibits the highest linear polarization of all runs. Quantitatively, we show the LP fraction and $I$ in Fig. \ref{fig: LP fraction (t) and I(t)} as a function of time, offset by a time $T_i$, such that the time-intervals in Tab. \ref{tab: Ray-tracing & GRMHD time intervals} have a common starting time for comparison purposes. The LP fraction peaks at $\sim 16\,\%$ for \texttt{K.99} compared to $\sim 11\,\%$ in the normal accretion states for both \texttt{K.80} and \texttt{K.90}. The LP-pattern itself of \texttt{K.99} differs as well, in that the EVPA longitudinal component (to the jet position angle) inside the jet is clearly dominant, as can be seen by the straight lines over a wide area inside the jet. Comparing this observation to \texttt{K.80} and \texttt{K.90}, those on the other hand, exhibit visible transverse components in most parts of the EVPAs. At the jet edge, this behaves analogously for \texttt{K.99}, where the EVPAs now sharply transition from following the jet longitudinally in the inside to a dominant EVPA-field perpendicular to the jet edge. This hints at a highly ordered magnetic field in the jet spine.
\\Notably, the jet structure of \texttt{K.99} is very distinct from the usual paraboloidal shape that is associated with MAD-simulations. Beyond the jet base, about $\sim 30\,\unit{\mu as}$ from the core, the jet transitions into a cone-like shape, which will be discussed in more detail in the next section.

\subsection{Jet edge and width profiles}

\begin{figure}
    \centering
    \includegraphics[width=\linewidth]{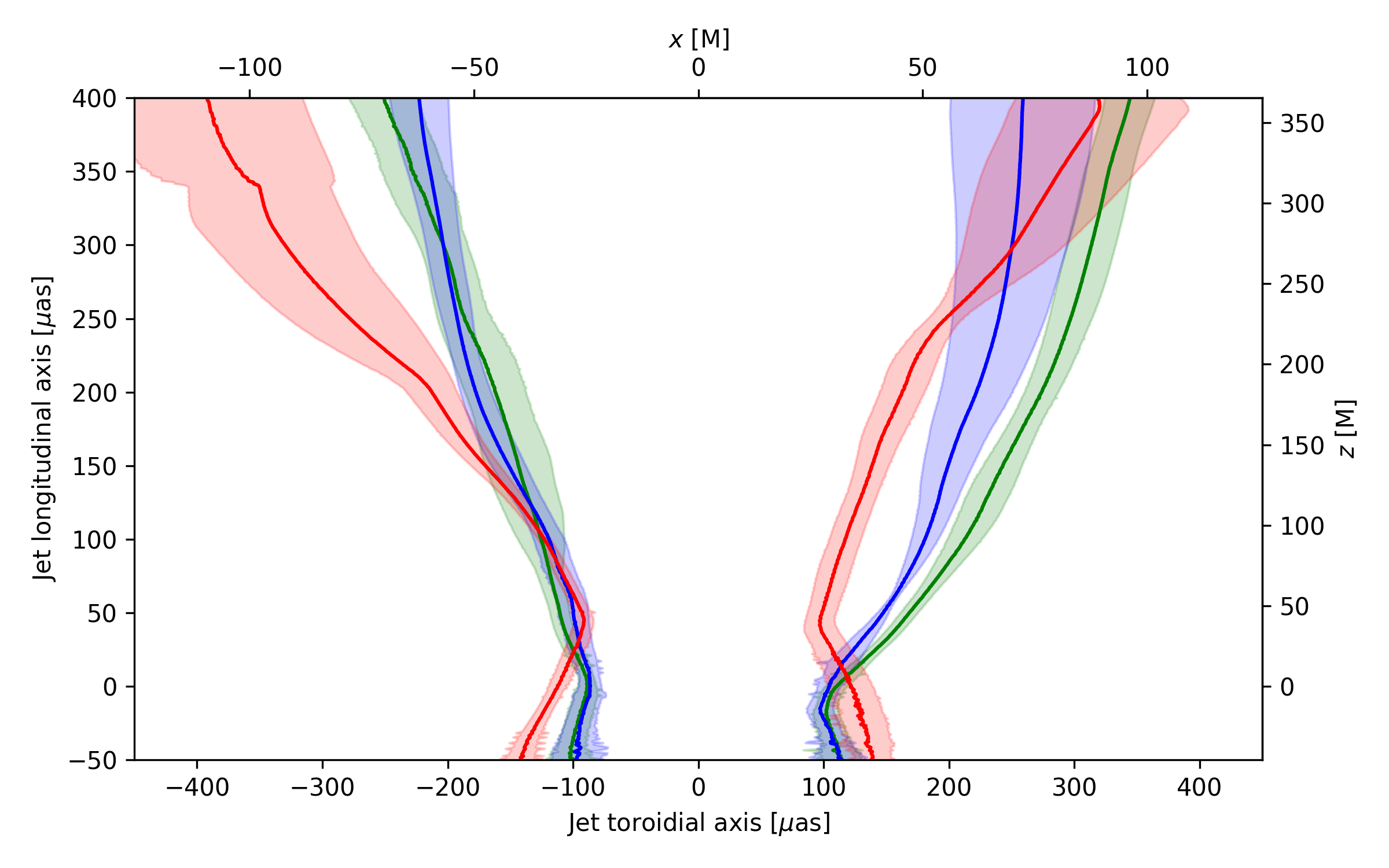}
    \caption{$\langle I\rangle = 5\times10^{-6}\,[\mathrm{Jy}]$ contours of time-averaged 86 GHz ray tracings (not the line integral convolved images shown in Fig. \ref{fig: time averaged ray-tracings}) for simulations \texttt{K.80} (green), \texttt{K.90} (blue), and \texttt{K.99} (red). The colored areas are $1\sigma$ bands of the standard deviations in the toroidal direction.}
    \label{fig: RT contours}
\end{figure}

In Fig. \ref{fig: RT contours}, we trace the profile of the jet analogously to the structural investigation of the jet sheath in section \ref{sec: GRMHD jet profiles}. We extract the time-averaged contours at $\langle I \rangle = 5\times10^{-6}\,\unit{Jy}$ (solid lines) and illustrate with the corresponding colored areas the variability perpendicular to the jet position angle (PA), i.e. the $1\sigma$-bands of the variation in jet-PA-perpendicular direction of the $I = 5\times10^{-6}\,\unit{Jy}$ contour over time. Therefore, the images were de-rotated to a jet PA of $-180^\circ$ (South-North direction) such that the jet can be subdivided into jet PA perpendicular slices for each pixel along the jet direction. The closest pixel to the contour is determined at each time step. Eventually, the mean and standard deviation at each slice over time can be calculated. The chosen contour of $I=5\times10^{-6}\,\unit{Jy}$ is a sensible tracer of something that could be considered the edge of the jet, which is in itself loosely defined, since it traces the apparent jet edge visible from Fig. \ref{fig: time averaged ray-tracings} at a dynamic range of about four orders of magnitude. In observations of the jet of M87* often the observed edge-brightening is the best indicator of the jet edge (often taken to be the HWHM towards the outside of the jet of the bright Gaussian edge features). In these simulations on most scales the jet perpendicular flux-profile does not exhibit bi- or multimodal distributions, especially not time-averaged, such that bright spots are not as pronounced on the edge of the jet.\\
The mean contours of \texttt{K.80} and \texttt{K.90} are very similar, whereas \texttt{K.99} stands out by its partially cone-like shape as pointed out in the last section. The actual difference in the asymmetries of the models becomes very apparent. In a basis defined by coordinates $(x,z)$ as in Fig. \ref{fig: RT contours}, we find that the asymmetry is strongly shifted to negative $x$-values in \texttt{K.99}. The variability of the jet structure provides confidence that the jet structures shown in Fig. \ref{fig: time averaged ray-tracings} are not distorted because of strong time-dependent morphological variations. The variation of these contours in $x$-direction slowly increase in $z$-direction in the cases of \texttt{K.80} and \texttt{K.90}, but increase rather abruptly and asymmetrically for \texttt{K.99} on the positive $x$ half-space at $\sim 250 \,\unit{M}$ and on the negative one at $\sim 200\,\unit{M}$. This can be explained by a helical arc propagating through the jet sheath, seen in Fig. \ref{fig: time averaged ray-tracings} downstream the jet in \texttt{K.99}. Nevertheless, the jet shape of \texttt{K.99} is completely altered in comparison to \texttt{K.80} and \texttt{K.90}.\\ \\

\begin{figure}
    \centering
    \includegraphics[width=\linewidth]{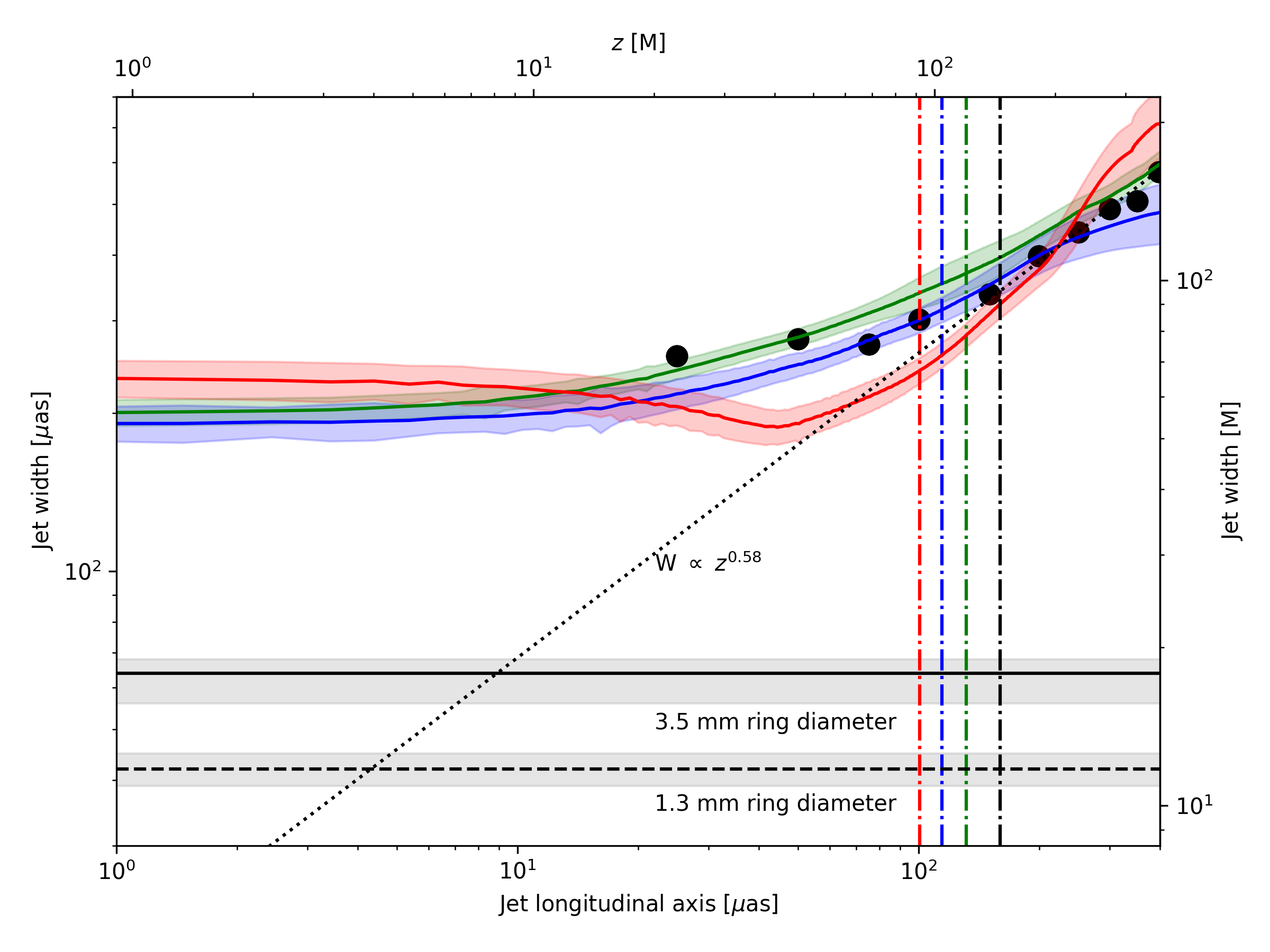}
    \caption{Jet width of M87 at 86 GHz along the longitudinal jet axis for simulations \texttt{K.80} (green), \texttt{K.90} (blue), and \texttt{K.99} (red) with corresponding $1\sigma$ bands. The black dots are data points of the jet width of M87 from the observation of \cite{Lu_2023}. The dotted line is a power-law fit from observations of M87. The solid horizontal line is the estimated 86 GHz ring size, and the dashed horizontal line is the 230 GHz ring size with respective error band (gray areas). The vertical lines are the distances from the BH, where the intrinsic half-opening angles equals the jet viewing angle of $\theta_\mathrm{v}=17^\circ$ (color-code is consistent).}
    \label{fig: jet width}
\end{figure}
\noindent
From these jet profiles of the $86$ GHz ray-tracing we can now extract the jet widths as a function $W(z)$ of the jet longitudinal axis $z$ (as defined before) and compare these to the observations of \cite{Lu_2023} shown in Fig. \ref{fig: jet width}. We note that the data-points of the jet-width measurements do not show the error-bars in \cite{Lu_2023}, hence we cannot elaborate on the statistical certainty of this comparison. From Fig. \ref{fig: jet width}, it is apparent that the theoretical predictions especially of \texttt{K.80} and \texttt{K.90} do match the observations well at the scales below $100\,\mu as$, contrarily to the claims of \cite{Lu_2023} that MAD simulations may have difficulty explaining the observed jet width at these scales. At the scales shown above $100\,\unit{\mu as}$ the simulations \texttt{K.80} and \texttt{K.90} converge to the extrapolated $W(z)\propto z^{0.58}$ power law fit from large scale observations (see \cite{2012Asada}, \cite{Hada_2013}, \cite{2016Hada}). \texttt{K.99} follows the power-law further up-stream and exhibits a sudden increase of the jet width at $\sim 40\,\unit{\mu as}$, whilst even closer to the BH it becomes almost constant, but even wider than the other models. This sudden increase in the width of the jet can be explained by the counter-jet, which exhibits a wider opening angle and thus shifts the jet edge of the counter-jet to positive $z$-values up to $\sim100\,\unit{M}$ (cf. Fig. \ref{fig: time averaged ray-tracings}, right image). For the common choice of the R-$\beta$ electron temperature- and $\kappa$ non-thermal electron prescriptions chosen here \texttt{K.99} exhibits a steeper slope than this observation suggests, rather following the large-scale power-law. The new observational data points at the scales closer to the jet launching region suggest that \texttt{K.90} satisfies these observations very well, as the data points are very well within the variability of \texttt{K.90} (cf. green band in Fig. \ref{fig: jet width}). We also see that this holds mostly for the common model \texttt{K.80}, as well.

\subsection{Ray-tracing the flux eruptions}

Since we mostly study the time-averaged quantities of the simulations presented in this work to extract general physical differences from the three models, we have to also shed light on the influence of the episodic magnetic flux eruptions on the radiative signatures (cf. in Sec. \ref{subsec: GRMHD flux eruptions} we discussed the consequences of the flux eruptions on the GRMHD quantities). In Fig. \ref{fig: K.99 Flux tube RT disk and jet} we show an example of a ray-tracing of such a flux eruption in the case of model \texttt{K.99}, where on the left we only ray-traced the disk emission and on the right the full emission that includes emission from the jet, as well. By comparison of both images, the influence of the flux eruption on the jet morphology becomes clear. The jet base strongly deviates from a paraboloid as the flux eruption envelopes the jet in a azimuthally asymmetric manner (cf. the feature in the upper half of the right image in Fig. \ref{fig: K.99 Flux tube RT disk and jet}). Since these flux eruptions are intermittent phenomena, they do not influence the radiative imprints on the jet morphology steadily and azimuthally symmetric, and are thus important to consider for the study of the long time-scale structural studies of MAD simulations. In order to be able to average out the strong azimuthal asymmetries introduced by MAD states, they have to be averaged over multiple ones. Since they occur quasi-periodically with a frequency of the order of $\sim 1000\,\unit{M}$, this would make studies on time-scales of at least $10000\,\unit{M}$ necessary. Hence, we limited our discussion to one MAD event per simulation.

\section{Conclusions}

We find that the initial magnetic energy content of the torus systematically alters the jet energetics and morphology of the developed simulations. It mostly affects the EM contribution in the jet, and from that it gains $\sim 60\%$ efficiency in the conversion from rest-mass energy to the energy of the outflow by means of increasing the magnetic energy of the torus at the initialization. This in turn can make nonthermal particle acceleration from vacuum gaps and subsequent pair cascades more likely, due to the increased EM field in the jet spine (see \cite{Levinson_2011}, \cite{Wendel_2021} for details on the vacuum-gap model and applications to low-luminosity AGNs such as M87 and Mrk501). We also find, that an increased magnetic filling of the FM torus elevates the outward angular-momentum transport, primarily in the jet, which is driven by Maxwell stresses. Comparing the angular-momentum transport of BHs with very high spin presented here to those of \cite{Chatterjee_2022}, in which the case of a non-rotating BH in a standard MAD-setup ($F=0.8$) was investigated, the major differences are the following: $i)$ the total radial angular-momentum flux is dominated by radially outward-directed flux in the jet and wind, mostly originating from the jet base; and $ii)$ morphologically, the angular-momentum vector fields are distinct, in that the high-spin BH creates field lines that point radially outwards in a polar angle range from $\sim (0-70)^\circ$ or $\sim (110-180)^\circ$. Toward the equatorial plane, the field lines transition from a monopole geometry to a quadrupole geometry in the disk and torus. The dominant, Maxwell-stress-induced outward polar flux is redirected to an equatorially inward, advectively driven flux. In contrast, the case of a non-rotating BH exhibits dominant, advectively driven inward angular-momentum flux in a larger portion of the disk and torus, whereas the outward-directed Maxwell-stress-induced flux is subdominant in the jet, such that the net angular momentum is transferred to the central BH. The outward angular momentum transport also covers a much smaller polar angle compared to simulations with high spin, contributing far less to the strength of outward winds. Our simulations of MADs with high-spin BHs demonstrate the effect of extracting the angular momentum of the BH to the angular momentum of the outflowing plasma in the jet and wind, as originally proposed by \cite{1977BlandfordZnajek}. We further find that there is a maximum of all net angular momentum fluxes inside the ergopsheric region of the BH, as opposed to the net-constant fluxes for MAD simulations with no spin. \\
It appears that the disk's vertical flux tubes created by the magnetically arrested disk affect the jet structure more significantly for high filling factors. From our sampled MAD states, we find a tendency for more spatially extended flux tubes that affect the jet morphology by introducing significant azimuthal asymmetries in the case of \texttt{K.99;} these propagate along the jet at least up to $200\,\unit{M}$. The jet's sheath structure is more stable on these scales; i.e., the jet structure exhibits lower variability in the jet-perpendicular direction. However, since we restricted the jet-morphology analysis to a time interval, in that a single flux eruption occurs in each simulation, further detailed, statistical analyses of the impact of the filling factor on the quasi-periodic flux eruptions could provide valuable insights into the observable footprints that they have. \\ \\
The radiative transport calculations also show that the jet's sheath structure is more stable and collimated up to $\sim 200\,\mu as,$ supported by the higher EM energy flux contribution in \texttt{K.99}. We find significantly elevated LP fractions of up to $\sim 15\%,$ in agreement with observations of the \cite{EHTVIII}. The jet width relation over the jet axial distance shows that GRRT calculations of GRMHD simulations of magnetically arrested accretion flow are able to match the observed jet width profile of \cite{Lu_2023} very well for the reference models. For \texttt{K.99,} we find deviations from the jet width, as it follows the observed large-scale power-law relation $W\propto z^{0.58}$ to shorter distances, whereas the observations follow a sub-power-law relation. The deviation of \texttt{K.99} from the observations, however, is a result of the different techniques used for the determination of the jet width. In the observations of \cite{Lu_2023}, the  HWHM of the outermost components perpendicular to the jet longitudinal axis determines the jet width, whereas we did not recreate the edge brightened jet morphology and thus used brightness contours to determine the jet width. This made the detailed jet--width relation difficult to compare to these observations; nevertheless, we show that all our MAD simulations provide a wide jet base substantially wider than the pure Blandford--Znajek (BZ) jet \cite[see][]{1977BlandfordZnajek} for a high-spin BH ($a\rightarrow1$). It is evident from our analysis of the jet energy fluxes that the almost evacuated, highly magnetized jet spine ($\sigma \gtrsim 1$) is dominated by EM energy flux, and in the enveloping jet sheath the wind contributes to the outflowing kinetic energy flux that increases with distance. This jet-sheath region ($\sigma \lesssim 1$ and the gravitationally unbound part of the plasma) is the origin of the jet emission, and the $\sigma=1$ contour is thus not a good tracer for the outermost jet edge that could be observed. We remind the reader that $\sigma=\sigma_\mathrm{cut}=1$ is the value for the emission cutoff for the jet spine, where we do not expect significant synchrotron emission in the radio regime because of the extremely low electron densities, as well as efficient synchrotron cooling, due to the presence of high magnetic-field strengths in $\sigma>>1$ regimes. We stress that a simple extrapolation and comparison of plasma parameters such as the magnetization from simulations to trace the edge or width of the jet observed at millimeter wavelengths \cite[see][Fig. 3; gray band shows $\sigma$ contours from force-free electrodynamic simulations of a BZ jet for different BH spins]{Lu_2023} is insufficient to approximate the complex emission processes, which are yet to be fully understood. From the GRRT of our GRMHD simulations for the jet of M87*, it is evident that the apparent width of the jet base is not anchored at the ergosphere of the BH, as opposed to force-free electrodynamic simulations. \\ \\
Overall, we find the jet of the \texttt{K.99}-model to exhibit more flux downstream of the jet compared to the reference models. This aids in solving the issue that GRMHD simulations produce jets that decline too much in flux downstream compared to observations. Hence, this model is well suited for modeling jets on larger scales, beyond the jet launching region. This is very helpful, since most SMBH systems in the Universe are of smaller angular size than M87*. With the enhanced jet power in this model, the jet remains brighter and more collimated on scales even beyond $1000\,\unit{M}$.\\ \\
The magnetic filling factor regulates the spatial extent of the initial vector potential and thus that of the magnetic field, as well as the total magnetic energy inside the initial torus configuration of our simulations, which are commonly used to model magnetically arrested accretion and also constitute the main portion of the simulation library used for M87 by the \citeauthor{EHTV}.
In general, we conclude from this study that the evolution of these MAD simulations is very sensitive to minor changes in the initial magnetic field in the FM torus, which was set by the filling factor. In order to keep track of the vast parameter space of MAD simulations and consequences of the initial conditions, the specification of the detailed initial vector potential of simulations is therefore paramount. \\ \\ \noindent
\small{ \textit{Acknowledgements.} We thank Charles F. Gammie for the helpful discussion regarding this work. This research is supported by the DFG research grant ``Jet physics on horizon scales and beyond" (Grant No.  443220636) within the DFG research unit ``Relativistic Jets in Active Galaxies" (FOR 5195). The numerical simulations and calculations have been performed on \texttt{MISTRAL} at the Chair of Astronomy at the JMU Wuerzburg. YM is supported by the National Key R\&D Program of China (grant no. 2023YFE0101200), the National Natural Science Foundation of China (grant no. 12273022, 12511540053), and the Shanghai municipality orientation program of basic research for international scientists (grant no. 22JC1410600)}
\bibliographystyle{aa_url.bst}
\bibliography{biblio}

\begin{appendix}
    \normalsize
    \section{Initial magnetic-field configuration and filling-scaling of the tori}\label{sec: Appendix Initial B-field}

    \begin{figure}[!h]
        \centering
        \includegraphics[width=.9\linewidth]{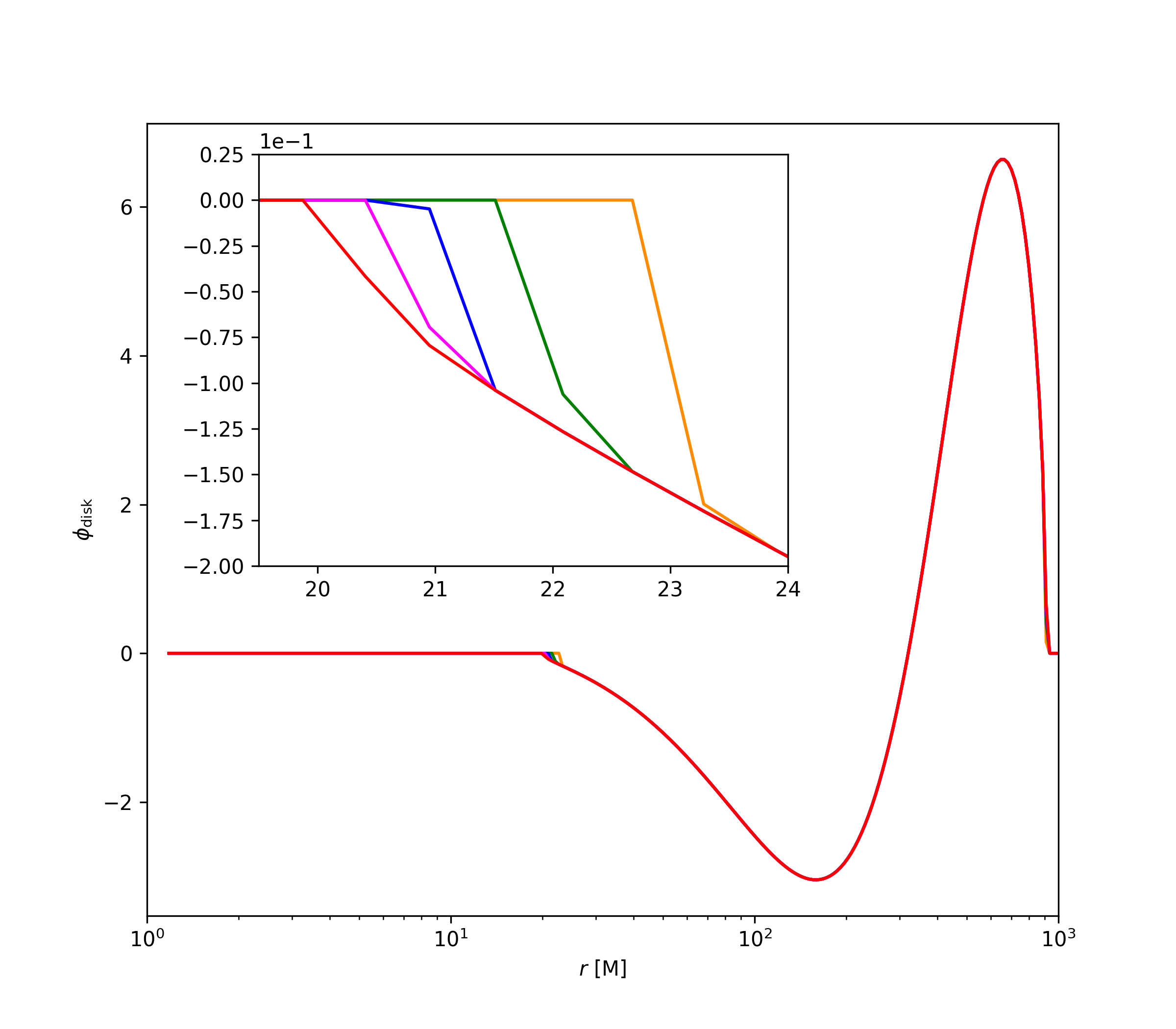}
        \caption{Initial disk-vertical magnetic flux permeating the disk. The colors correspond to simulations \texttt{K.60} (orange), \texttt{K.80} (green), \texttt{K.90} (blue), \texttt{K.95} (magenta) and \texttt{K.99} (red).}
        \label{fig: initial disk flux}
    \end{figure}

    \noindent
    Here, we calculate the initial covariant volume fractions of the volume containing the cells in which the initial invariant magnetic-field strength $b$ is defined and the initial volume enclosing the full mass density distribution of the torus. The initial magnetic field is normalized such that the minimum of plasma $\beta$ in the torus is $\beta_{\rm{min}}=100$ in all simulations. We calculate the total and shell-integrated covariant volumes of the torus and enclosed magnetic-field as follows:
    \begin{align}
         & V_\mathrm{t} = \int_{\{rho>8\times10^{-7}\}} \sqrt{-g}\,dr\,d\theta\,d\phi, \\
         & V_\mathrm{b} = \int_{\{b>1\times10^{-7}\}} \sqrt{-g}\,dr\,d\theta\,d\phi.
    \end{align}
    \noindent
    We chose the masking conditions for the enclosed volume of the density distribution such that $\rho$ is above the density distribution in the atmosphere (the space excluding the torus and BH), that should have its minimum at $\rho_\mathrm{min}=10^{-6}$ (as stated in Sec. \ref{sec: GRMHD}), however, increases up to $\rho \sim 8\times10^{-7}$ in the vicinity of the BH. For $b$, the condition is not as strict, since its numeric value is only defined, where the four-vector potential $A_{\phi}$ contributes to the curl to generate the magnetic field. We choose a low threshold of $b>1\times10^{-7}$ and set the non-defined values that are mostly outside the torus to $0$ in order to capture the full volume of the initial magnetic field. The actual filling factor is then given by:
    \begin{equation}
        f_\mathrm{a} = \frac{V_\mathrm{b}}{V_\mathrm{t}}.
    \end{equation}
    \noindent
    We note that, $f_{\rm{a}}$ is a measure for the volumetric magnetic-field extent, or the magnetic filling of the initial magnetospheric configuration, that we calculate for each filling factor $F$ and is not equivalent to the latter values as can be seen in Tab. \ref{tab: actual torus fillings}.
    \begin{table}[!h]
        \centering
        \caption{Initial volume percentage of the torus filled by magnetic fields $f_\mathrm{a}$ compared to the initial parameters $F$, the filling factors, for each simulation.}
        \begin{tabular}{c|c|c|c|c|c}
            \hline
            Simulation          & \texttt{K.60} & \texttt{K.80} & \texttt{K.90} & \texttt{K.95} & \texttt{K.99} \\ \hline
            $F$                 & 0.60          & 0.80          & 0.90          & 0.95          & 0.99          \\ \hline
            $f_\mathrm{a}$ [\%] & 99.16         & 102.06        & 103.57        & 104.41        & 105.62        \\ \hline
        \end{tabular}\label{tab: actual torus fillings}
    \end{table}
    \noindent
    The total volume containing magnetic fields is generally quite close to the constant torus volumes throughout the simulations, much closer than the filling factors $F$ suggest. However $F$, simply scales the offset of the initial vector potential $A_{\phi}$ (cf. Eq. \eqref{eq: initial vector potential}), essentially increasing its offset from the zero-cutoff maximum function in Eq. \eqref{eq: initial vector potential} as can also be seen in Fig. \ref{fig: initial A+B-field}. Still, increasing $F$ indeed leads to an increased volume that contains magnetic fields. Hence, we use the filling factor synonymously with the scaling of the magnetic-field extent and the magnetic energy in the FM-torus.

    \section{Average MHD parameters}\label{sec: Appendix MHD parameters}

    \begin{figure}[!h]
        \centering
        \includegraphics[width=.9\linewidth]{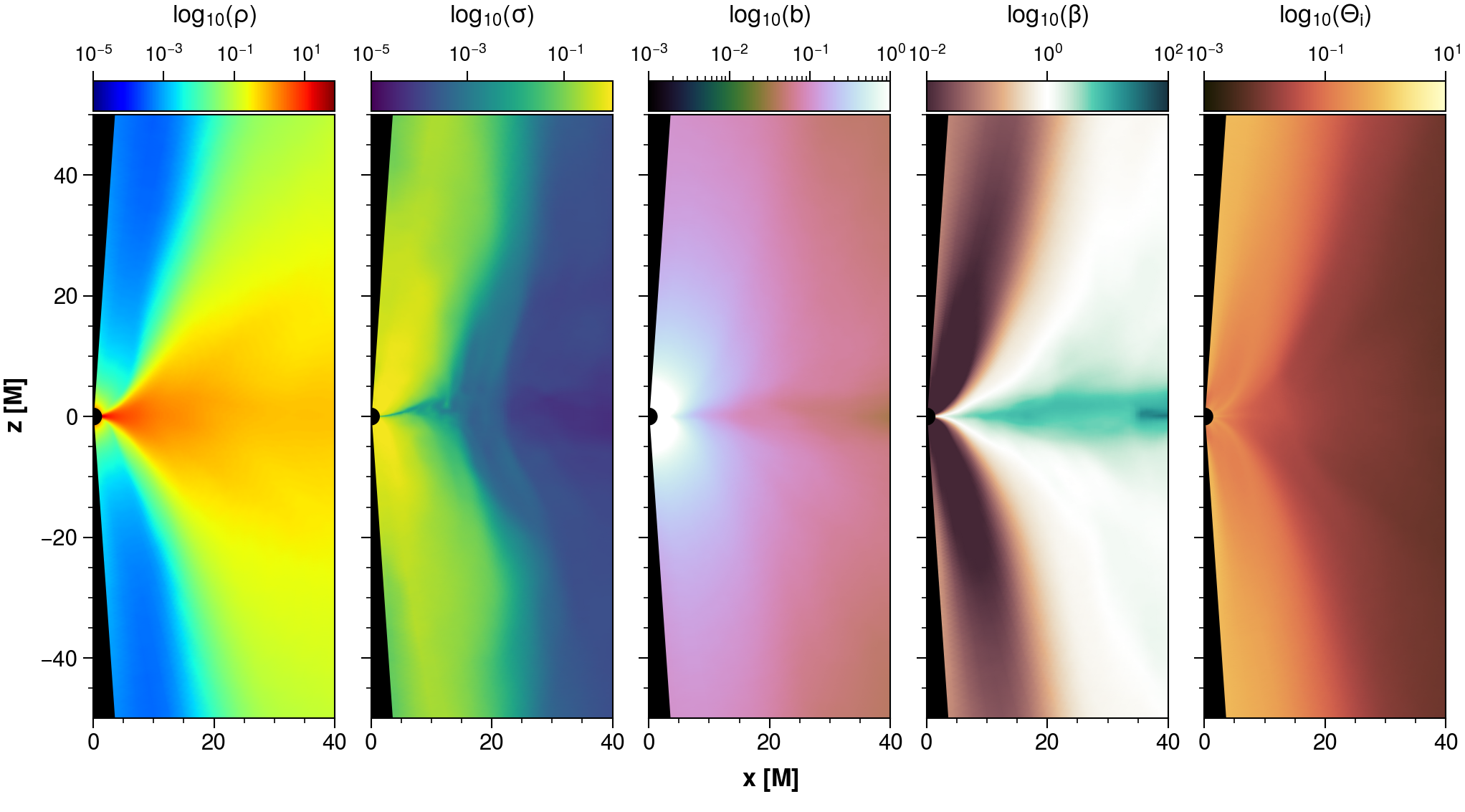}
        \caption{Time and azimuthally averaged plasma (MHD) parameters of model \texttt{K.99}. The time is averaged over the interval indicated in Fig. \ref{fig: comparison interval} and listed in Tab. \ref{tab: Ray-tracing & GRMHD time intervals}. We apply a polar cut of $\theta_\mathrm{cut}=4^\circ$ in the post-processed ray tracings for all models, which is indicated by the black regions at the poles.}
        \label{fig: time and azimuthal average plasma K.99}
    \end{figure}

    \section{Mass-accretion rate}

    \begin{figure}[h!]
        \centering
        \includegraphics[width=.85\linewidth]{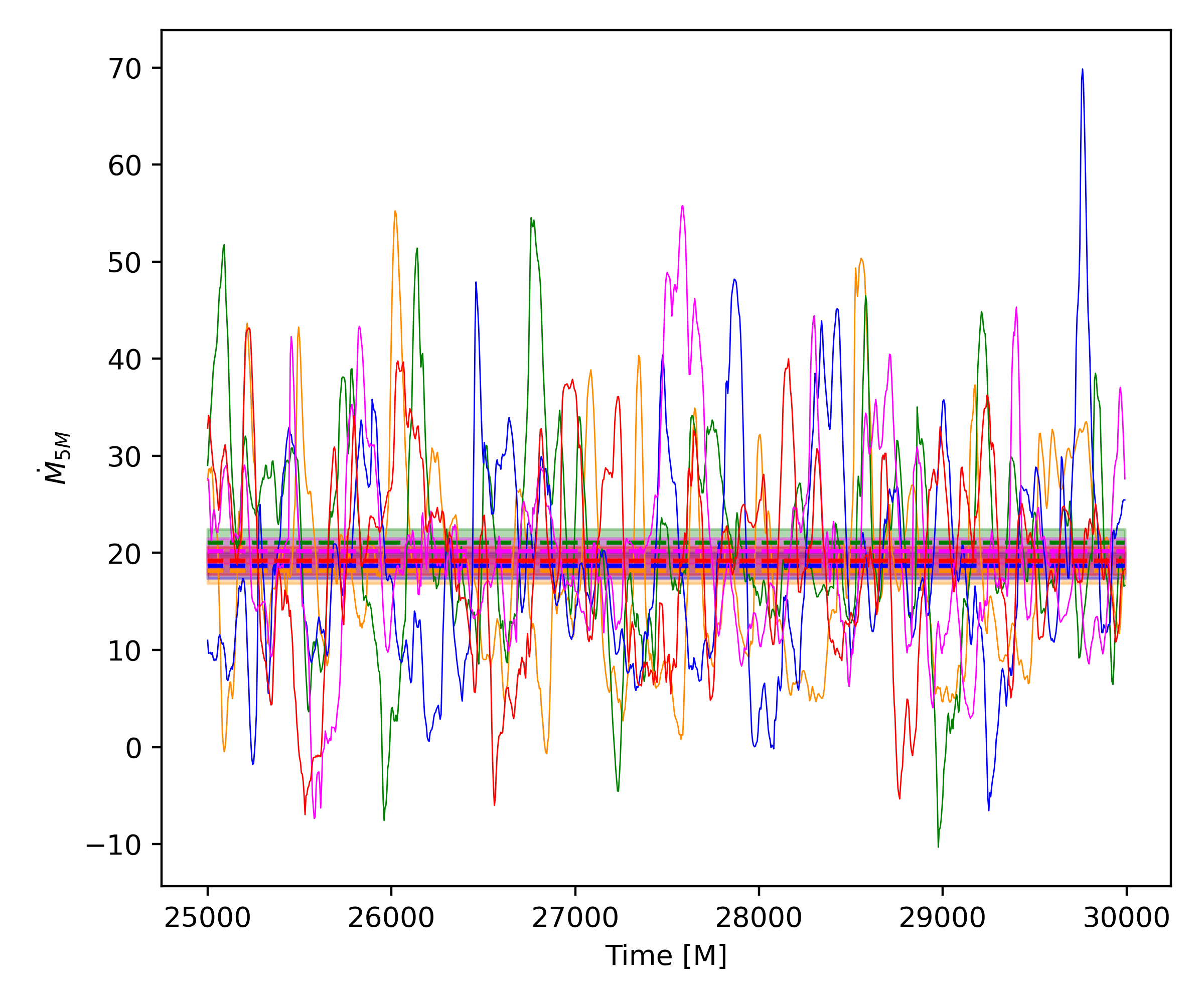}
        \caption{Mass-accretion rate $\dot{M}_{\rm{5M}}$ of simulations \texttt{K.60} (orange), \texttt{K.80} (green), \texttt{K.90} (blue), \texttt{K.95} (magenta) and \texttt{K.99} (red). The horizontal lines are the averages and the shaded area span the $2\sigma_{\rm{corr}}$ confidence interval of the correlated time series. The overlapping error-bands show that the mass-accretion rate is sufficiently similar in between all simulations, thus, indicating that these are in a steady-state. }\label{fig: Mdot 5M comparison}
    \end{figure}

    \newpage
    \onecolumn
    \section{MRI quality factors}\label{sec: Appendix Q-factors}

    \begin{figure*}[!h]
        \centering
        \includegraphics[width=\linewidth]{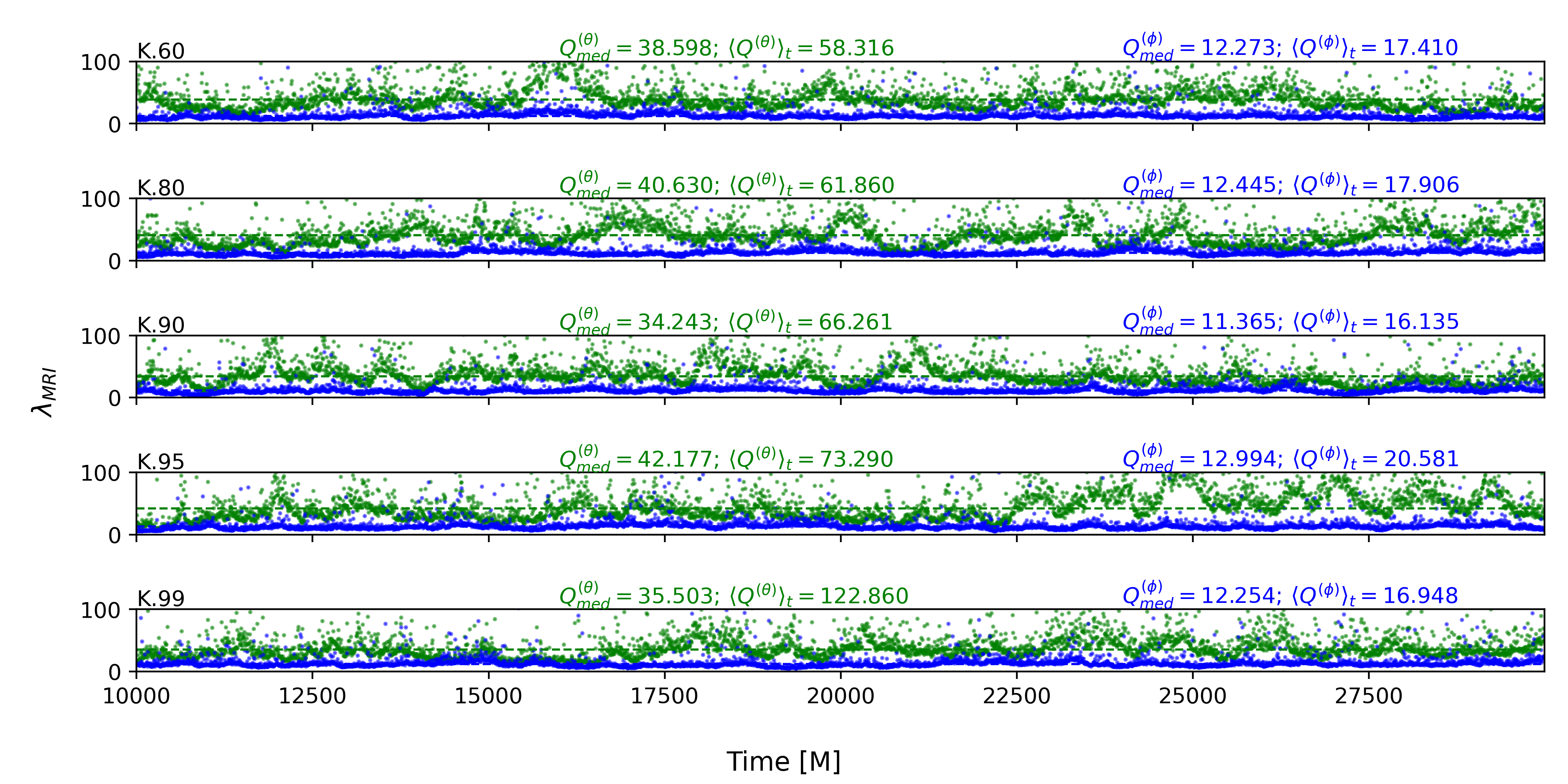}
        \caption{Disk averaged MRI quality factors in $\theta$ and $\phi$ of each simulation, row-wise. Dashed lines are the median values taken over the time span shown in the graphs. Besides the median values we also give corresponding time averaged values. Due to the skewed nature of the temporal distriubtion of the disk averaged $Q$ factors, the median is more appropriate for interpretation. It is noted that for $Q^{(\theta)}$ $\sim (3-6)\%$ of points lie outside the graph, i.e. are above $100$ in the three simulations. For $Q^{(\phi)} \sim 1\%$ are concerned. }
        \label{fig: Q-factors temporal}
    \end{figure*}

    \begin{figure}[!h]
        \centering
        \includegraphics[width=.5\linewidth]{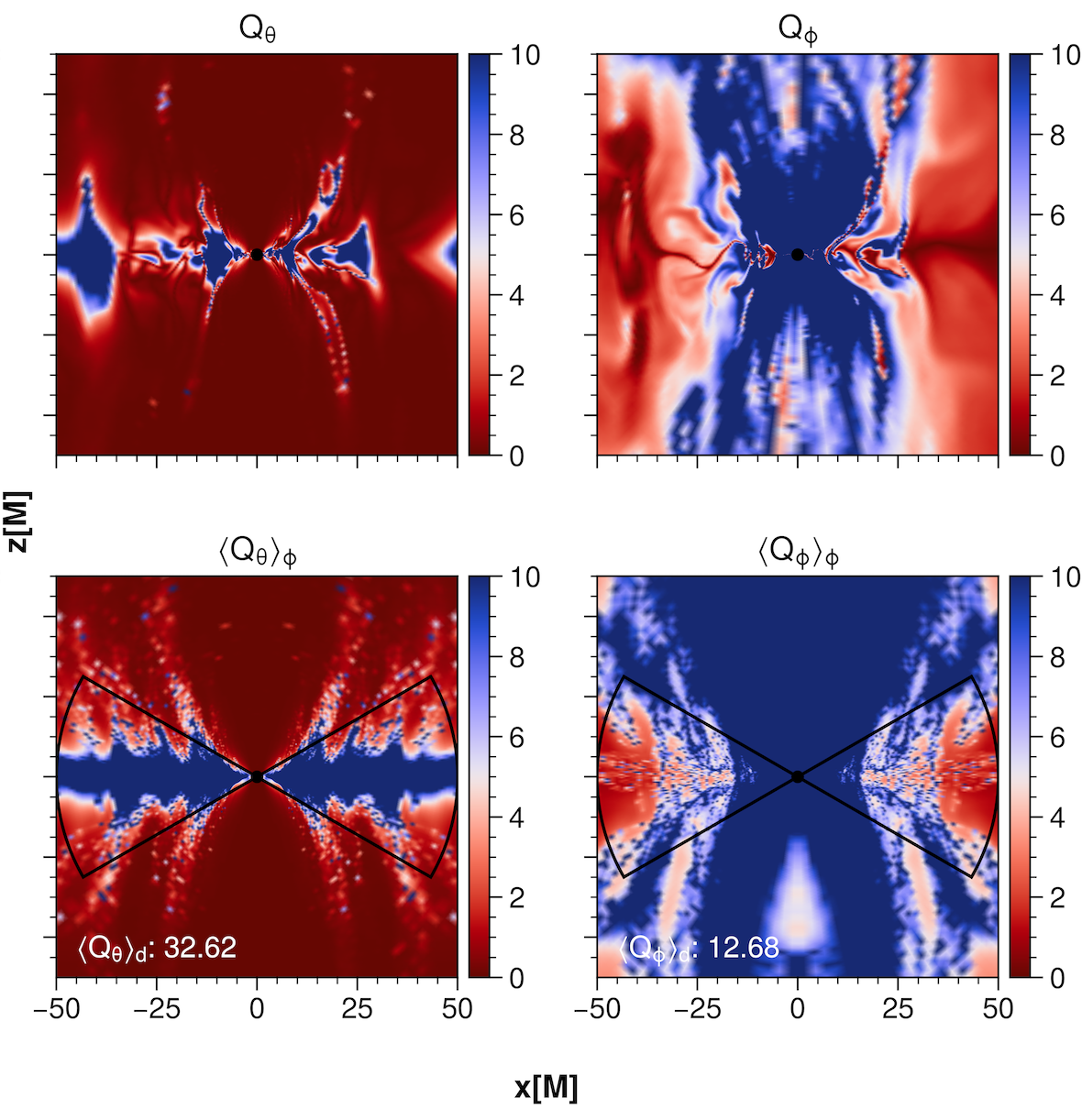}
        \caption{Meridional cross-section of MRI Quality factors in $\theta$ and $\phi$ directions. Top row: Meridional cross-section. Bottom row: Volume averaged $Q$ factors over the full azimuthal angle. The black lines indicate the regions where the disk is cut, poloidally, and subsequently averaged over the full azimuthal angle to obtain the disk averaged values given in the bottom left corners.}
        \label{fig: Q-factors snapshot and azimuthal average}
    \end{figure}

    \twocolumn
    \section{Error estimation of time series}\label{Appendix Error Estimation}

    Here we elaborate on the error estimation of the correlated time series, inspired by \cite{lalakos2025}. The reduction of the number of independent samples can be quantified by the integrated autocorrelation time $\tau_{\rm{int}}$ of a time series $f(t)$, which is defined as:
    \begin{equation}
        \tau_{\rm{auto}} = \left (1 + 2\sum_{k=1}^{N} \rho(k\,\Delta t) \right ) \Delta t \,,
    \end{equation}
    \noindent
    where $\rho(t)$ is the normalized autocorrelation function of the unity normalized time series $f(t)$ at lag $t$, $N$ is the number of samples and $\Delta t$ is the sampling interval (5M in our case) for normalization. The normalized autocorrelation function is defined as:
    \begin{equation}
        \rho(k\, \Delta t) = \frac{1}{\sigma^2} \sum_{i=1}^{N- k\, \Delta t} (f(i\,\Delta t) - \mu)(f(i\, \Delta t + k \Delta t) - \mu).
    \end{equation}
    The autocorrelation function is computed up to a lag of $N/2$ to avoid noise-dominated estimates at large lags and normalized to lie between values of $-1$ and $1$ to enable comparability between the simulations.\\
    The error of the mean $\mu$ of a correlated time series of duration $T$ is then given by:
    \begin{equation}
        \sigma_{\rm{corr}} = \sigma \sqrt{\frac{2\tau_{\rm{auto}}}{N}} \,,
    \end{equation}
    \noindent
    where $\sigma$ is the standard deviation of the time series.
    We typically found values of $\tau_{\rm{auto}} \lesssim 10 M$ corresponding to an error $\lesssim 5$ higher than the standard error of the mean, when assuming the data are entirely uncorrelated. Therefore, we avoid underestimating the uncertainty of the mean.\\

    \onecolumn
    \section{Angular momentum fluxes at varying radii}

    \begin{figure*}[!h]
        \centering
        \includegraphics[width=\linewidth]{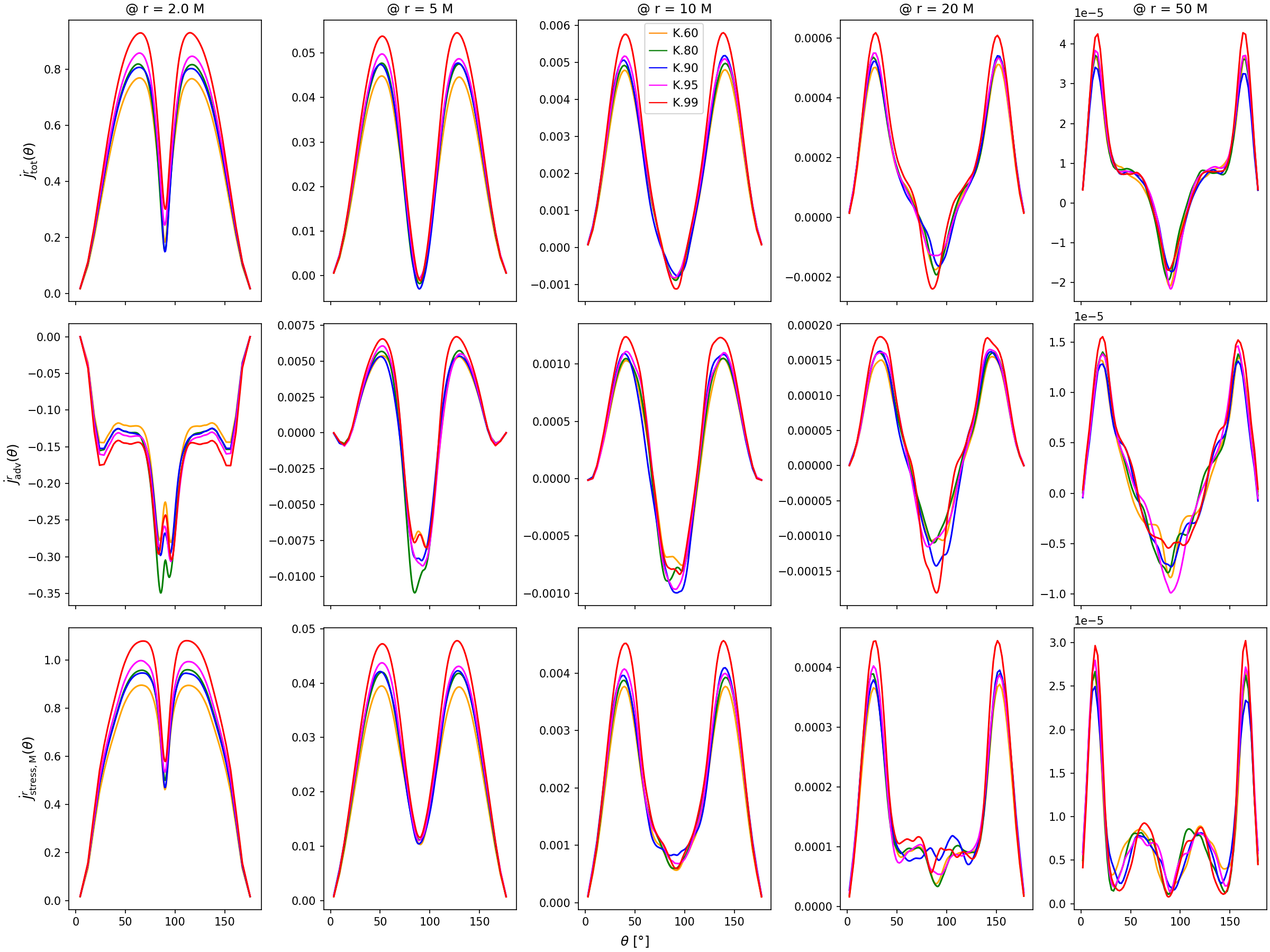}
        \caption{Polar slices of the radial angular momentum flux at varying radii increasing to the right. Solid lines are the radially outward flux and dashed lines are inward flux. Note the varying scales on the y-axis.}\label{fig: polar slices angularmomentum fluxes}
    \end{figure*}

    \section{Morphology of a flux eruption snapshot}

    \begin{figure}[!h]
        \centering
        \includegraphics[width=.7\linewidth]{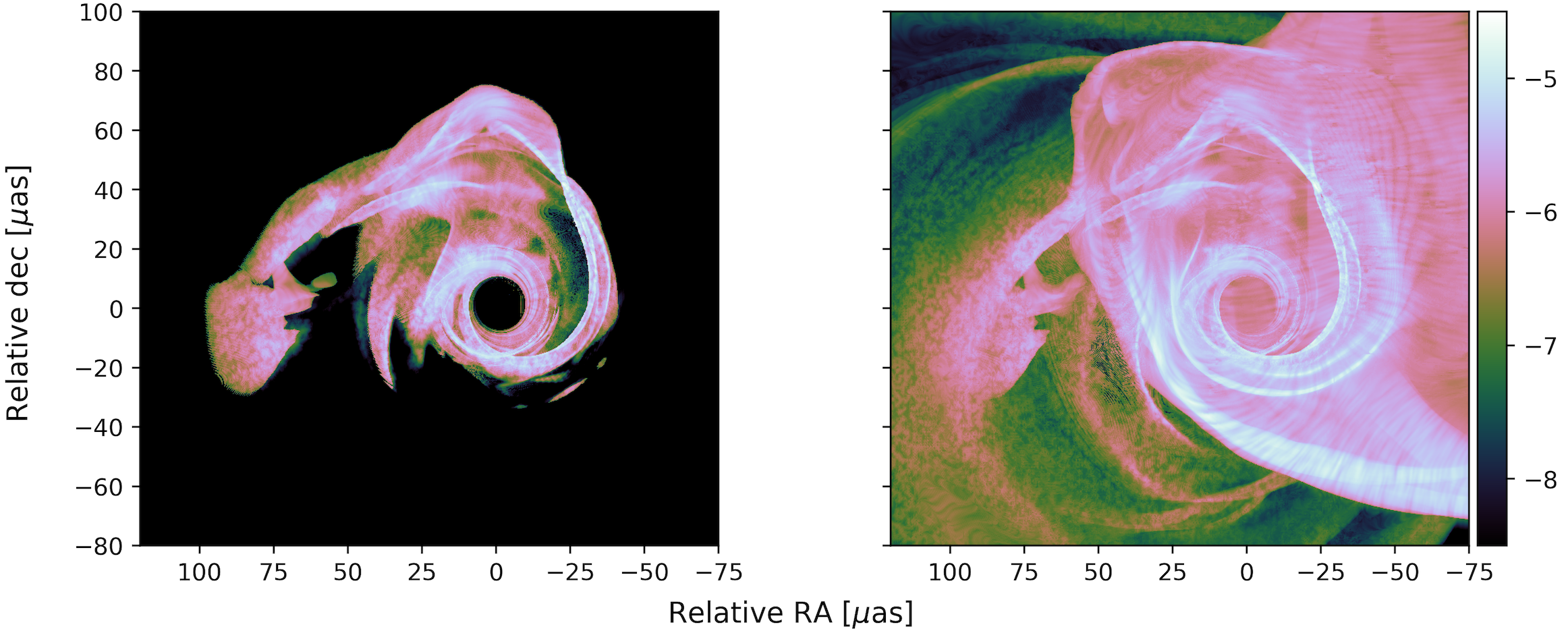}
        \caption{86 GHz ray-tracings of the flux eruption of the \texttt{K.99} run (same procedure as in Fig. \ref{fig:  time averaged ray-tracings} for the line integral convolution). Left image: only ray-traced thermal disk emission. Right image: ray-traced non-thermal jet and thermal disk emission}\label{fig: K.99 Flux tube RT disk and jet}
    \end{figure}
\end{appendix}

\end{document}